\documentclass[twocolumn,aps,prc,superscriptaddress]{revtex4-2}

\usepackage{color,amsmath,amssymb,epsfig,graphicx}
\usepackage{bm}
\usepackage{mathptmx, courier, pifont}
\usepackage[scaled=0.92]{helvet}
\usepackage[T1]{fontenc}
\usepackage{textcomp}
\usepackage[colorlinks=true,linkcolor=blue,filecolor=blue,urlcolor=blue,citecolor=blue]{hyperref}

\usepackage{multirow}
\usepackage{longtable}
\usepackage[utf8]{inputenc}
 \def\ket{\!>\,} \def\ack{\,|\,}
\begin{document}
\author{W. Tawseef}
\affiliation{Department of Physics, University of Kashmir, Srinagar,
  Jammu and Kashmir, 190 006, India}

\author{Nazira Nazir}
\email{naziranazir238@gmail.com}
\affiliation{Department of Physics, University of Kashmir, Srinagar,
  Jammu and Kashmir, 190 006, India}

 \author{S. Jehangir}
 \email{sheikhahmad.phy@gmail.com}
 \affiliation{Department of Physics, Govt. Degree College Kulgam,
 Jammu and Kashmir, 192231, India}

\author{J. A. Sheikh}
\email{sjaphysics@gmail.com}
\affiliation{Department of Physics, University of Kashmir, Srinagar,
  Jammu and Kashmir, 190 006, India}
  
\author{C. Majumder}
\affiliation{Department of Physics, Indian Institute of Technology Bombay, Mumbai, 400076, India}

\author{S. Chakraborty}
\affiliation{Institute of Engineering \& Management, University of Engineering and Management, Kolkata, 700091, India}

\author{G. B. Vakil}
\affiliation{Department of Physics, University of Kashmir, Srinagar,
  Jammu and Kashmir, 190 006, India}

\author{G. H. Bhat}
\affiliation{Department of Physics, GDC Shopian, Higher Education, Jammu and Kashmir, 192 303, India}

\author{N. A. Rather}
\affiliation{Department of Physics, Islamic University of Science and Technology, 
  Jammu and Kashmir, 192 122, India}

%\author{S. Frauendorf}
%\email{Stefan.G.Frauendorf.1@nd.edu}
%\affiliation{Department of Physics, University of Notre Dame, Notre Dame, USA} 

%\title{ The fermionic underpinning of gamma softness in $\textsuperscript{104}$Ru 
%}
\title{ Microscopic investigation of magnetic and antimagnetic rotational motion in atomic nuclei }

\begin{abstract}

  In the present work, we have generalized the projected shell model (PSM) approach to include the quasiparticle
  excitations from two major oscillator shells, and have also extended the basis space to five-quasiparticle
  configurations for odd-mass nuclei. The magnetic and antimagnetic rotational structures observed in odd-neutron Pd- and Cd-isotopes
  have been investigated as a first major application of the new development. It is shown that PSM approach provides
  a reasonable description of the observed properties of magnetic and antimagnetic rotational bands.  
\end{abstract}

\date{\today}

\maketitle
\section{Introduction}
\label{Sect.01}

Atomic nucleus is among a few quantum many-body systems that displays a rich diversity of excitation modes which include
rotational, vibrational and single-particle mechanisms \cite{BMII,RS80}. The nuclei in the vicinity of the spherical shell
closures, generally, depict single-particle and vibrational excitation modes in the low-spin region, and nuclei far
from closed shells exhibit rotational motion. The rotational spectra have the characteristic feature of strong
electric quadruple transitions and can be understood by considering the collective motion of all the nucleons in the nucleus. In the mean-field
approach, a deformed potential with ellipsoidal shape describes the observed strong quadruple transitions \cite{BMII}. The
deformed potential defines the orientation with respect to the space fixed axis in an analogous manner to the diatomic molecules.
In most of the nuclei, the band structures can be generated by considering rotation about an axis perpendicular
to the symmetry axis of the spheroidal potential. This geometry leads to the signature quantum number with
dominant quadrupole transitions along each signature branch \cite{sf01,sf97}.

The above simple picture of the nuclear excitation modes prevailed
until a new class of band structures were discovered in the vicinity of the closed shell structures \cite{lm96,clark2000shears}.
It was observed that nuclei with a few proton holes and neutron particles in the Pb-region have rotational like band structures with
magnetic dipole B(M1) transitions dominating rather than electric quadrupole B(E2) transitions \cite{FRA2}. These nuclei
have very small deformation and the rotational collective
motion of a deformed system can be ruled out. The new band structures were interpreted \cite{FRA2,RMClark_1999,csm,PhysRevLett.87.132503}
in terms of ``shears'' mechanism with
proton holes and neutron particles forming two blades of a pair of shears. The angular-momentum of the proton holes, occupying
high-K orbitals, is oriented towards the symmetry axis, whereas the angular-momentum of the neutron particles, filling low-K orbitals,
is aligned towards an axis perpendicular to the symmetry axis. The high-spin states are then generated by closing the blades of 
shears with the highest angular-momentum obtained when the proton and neutron vectors coincide. The perpendicular
components of the neutron and proton dipole moments add up at the band head that results into a large B(M1) transition. However, the closing
of the blades with increasing spin leads to decrease in the dipole moment and consequently the B(M1) transitions are also reduced with
spin. The reduction of the B(M1) transitions have been observed in many systems and affirms the shears mechanism interpretation
of the rotational like band structures observed in the vicinity of the closed shells \cite{RMClark_1999,PhysRevLett.78.1868,clark2000shears,HUBEL20051,PhysRevC.78.024313}. The name magnetic rotation (MR)
has been ascribed to this
new phenomenon as it is the rotation of the magnetic dipole vector in space that is responsible in generating the band structures. 

It was predicted, soon after the MR bands were discovered, that it should be also possible to observe the antimagnetic rotation (AMR)
phenomenon \cite{sf97,sf01,FRau97,PhysRevC.99.041301,dft1,Wang2020In109AMR,MA2021122319}. In this alternative geometry, two neutron-proton pairs form subsystems of
two shears with the magnetic dipole vectors of the
shears anti-aligned. This arrangement has net zero dipole moment and is quite similar to anti-ferromagnetism in condensed matter physics.
In an antiferromagnetic material, one-half of the dipole moments in one sublattice is aligned in one direction and the other half is
oriented in the opposite direction in the second sublattice with net zero total dipole moment. At the band head, the angular momentum of the
proton holes is aligned towards the opposite sides of the symmetry axis, and angular-momentum of the neutron particles is directed along the
rotational axis. The AMR band is generating by simultaneously moving the two proton blades towards the angular-momentum vector of
the neutrons.

There are two characteristics features of the AMR phenomenon, one is the absence of the B(M1) transitions as the magnetic dipoles
of the two shears are anti-aligned. The second is that the system in the AMR configuration is symmetric with respect to rotation by $\pi$
about the total angular-momentum axis. This corresponds to the signature symmetry and the AMR band should consist of regular sequence of energy levels
differing in spin by $2\hbar$. The in-band E2 cascades should be weak considering that AMR configurations are weakly deformed.

The MR structures have been observed in several regions of the Segre chart and 225 bands observed in 114 nuclides are tabulated
in Ref.~\cite{kumar23}. The AMR, on the other hand, is a subtle phenomenon and, so far, 37 such  band structures have been reported in 27 nuclei \cite{kumar23}. It
was predicted \cite{sf97,sf01} as a alternative mechanism of generating angular-momentum and should occur in the same regions where
the MR mode have been observed. Nuclei in the A $ \sim $ 110 region, in particular the Cadmium isotopes, were proposed to be the
ideal candidates to explore the AMR mode \cite{Cd94,chiara20}. Cadmium has two proton holes in high-$\Omega$ orbitals in $1h_{9/2}$ subshell that will couple
with neutrons in low-$\Omega$ orbitals $1h_{11/2}$ to form the 
twin-shears configuration of the AMR mode. As a matter of fact, the first AMR band structure was identified in $^{109}$Cd \cite{chiara20,109cd} with lifetime
measurements clearly demonstrating a decreasing trend in the B(E2) values with increasing spin, and the ratio of the moment of inertia to B(E2)
of about 165 $\hbar^2$ MeV$^{-1}$ (eb)$^{-2}$.  AMR band structures have been reported in other nuclei, in particular, detailed investigations have been
performed for $^{101}$Pd \cite{101pdsuga,101pdsuga2,101pd}, $^{103}$Pd \cite{103pdamr}, $^{105}$Cd \cite{105cdamr} and $^{107}$Cd \cite{107cdamr} nuclides.  

Theoretically, several models have been proposed to elucidate the properties of MR and  AMR band structures. These include cranked shell model (CSM)
\cite{csm}, semiclassical model (SCM) \cite{chiara20,101pdsuga2,101pdsuga,109cd} and relativistic density functional theory (RDFT)
approaches \cite{dft1,dft2}. The CSM approach, as a matter of fact,
was originally used to predict the existence of the AMR mode \cite{FRau97}. The SCM approach is a semiclassical version of the particle-rotor
model with predefined shears arrangement for the valence nucleons. This model has been shown to provide a good description of the properties
of AMR bands. RDFT model has been used to investigate the high-spin properties of $^{105}$Cd and it has been shown that polarization effects
play an important role to describe the AMR features. 

In the present work, we have developed an alternative approach of the projected shell model (PSM)
to investigate the properties of the AMR and MR band structures. In the PSM approach, the angular-momentum projection is carried out, and the wavefunctions
have well defined angular-momentum. In comparison to the cranking approximation employed in the other models, the PSM approach is well suited to
study the electromagnetic transition probabilities. In the earlier version of the PSM approach, the valence particles were restricted to one
major oscillator shell \cite{KY95,SUN19941,PALIT2001141,SINGH201641}. In the present work, we have extended the model space and the valence nucleons  can occupy two major
oscillator shells. Further, we have also generalized the PSM approach to include up to five quasiparticle configurations in odd-mass
nuclei.
This extension, as will be discussed in the next section, is needed to explore the properties of the MR and AMR bands observed
in A $ \sim$ 110 region. The manuscript is organised in the following manner. In the next section, the extended PSM approach is presented
with a focus on odd-neutron systems. We have also employed the semiclassical particle rotor model to obtain the properties of the
Pd- and Cd- nuclides as an alternative approach. Although this approach is well documented \cite{Macchiavelli,macch1,144dy}, but for
completess, we  shall briefly discuss it in section~\ref{scm}. In section~\ref{result}, the results obtained on excitation energies, alignments and transition probabilities for chains of
odd-neutron Cadmium and Palladium nuclides are presented and discussed. Finally, the present work is summarized and concluded in section~\ref{sum}.

\section{Extended PSM approach}
\label{Sect.02}

In the present work, we have considerably extended the basis space in the PSM approach, and the complete set of configurations is given by $:$
\begin{equation}
\begin{array}{r}
%\hat P^I_{MK}\ack\Phi\ket;\\
~~\hat P^I_{MK}~ a^\dagger_{\nu_1} \ack\Phi\ket;\\
~~\hat P^I_{MK}~a^\dagger_{\nu_1}a^\dagger_{\pi_1}a^\dagger_{\pi_2}  \ack\Phi\ket;\\
  ~~\hat P^I_{MK}~a^\dagger_{\nu_1}a^\dagger_{\nu^{\prime}_2}a^\dagger_{\nu_3^{\prime}}  \ack\Phi\ket;\\
    ~~\hat P^I_{MK}~a^\dagger_{\nu_1}a^\dagger_{\nu_2}a^\dagger_{\nu_3}  \ack\Phi\ket;\\
  ~~\hat P^I_{MK}~a^\dagger_{\nu_1}a^\dagger_{\nu^{\prime}_2} a^\dagger_{\nu_3^{\prime}}a^\dagger_{\pi_1} a^\dagger_{\pi_2} \ack\Phi\ket;\\
  ~~\hat P^I_{MK}~a^\dagger_{\nu_1}a^\dagger_{\nu_2} a^\dagger_{\nu_3}a^\dagger_{\pi_1} a^\dagger_{\pi_2} \ack\Phi\ket,
\label{basis}
\end{array}
\end{equation}
where $\ack\Phi\ket$ is the  deformed quasiparticle vacuum state, and
$\hat P^I_{MK}$ is the 
angular-momentum-projection operator given by \cite{RS80} $:$ 
\begin{equation}
\hat P ^{I}_{MK}= \frac{2I+1}{8\pi^2}\int d\Omega\, D^{I}_{MK}
(\Omega)\,\hat R(\Omega),
\label{projection}
\end{equation}
with the rotation operator 
\begin{equation}
\hat R(\Omega)= e^{-i\alpha \hat J_z}e^{-i\beta \hat J_y}
e^{-i\gamma \hat J_z}.\label{rotop}
\end{equation}
Here, $''\Omega''$ represents the set of Euler angles 
($\alpha, \gamma = [0,2\pi],\, \beta= [0, \pi]$). In the present work axial symmetry has been assumed and the basis
configurations in Eq.~(\ref{basis}) have well defined ``K'' (projection of the angular-momentum along the symmetry axis)
value. This symmetry has been imposed as magnetic and antimagnetic rotational bands are considered to preserve
the axial symmetry. The advantage is that the three-dimensional integration reduces to only one as the intrinsic state
is an eigenstate of the $\hat{J}_z$ operator. 
%and   $\hat{J}^{,}s$ are the angular-momentum operators.

In Eq.~(\ref{basis}), $\nu$ and $\nu^{\prime}$ denote the quantities in two different oscillator shells. For instance, in the mass A $\sim$ 110 region, neutrons (protons) occupy N = 3,4,5 (2,3,4) shells, the index $\nu$ and $\nu^{\prime}$ represent neutrons occupying N = 5 and 4 shell respectively.  In the original version of the PSM program \cite{KY95}, particles are restricted to occupy only one
major oscillator shell \cite{KY95}.  

The above basis states are used to diagonalize the shell model Hamiltonian, given by $:$
\begin{eqnarray}
\hat H =  \hat H_0 -   {1 \over 2} \chi \sum_\mu \hat Q^\dagger_\mu
\hat Q^{}_\mu - G_M \hat P^\dagger \hat P - G_Q \sum_\mu \hat
P^\dagger_\mu\hat P^{}_\mu . \label{hamham}
\end{eqnarray}
The above Hamiltonian, consisting of pairing and quadrupole-quadrupole interactions, is the same as used in all earlier
PSM calculations \cite{KY95}
In the above equation, $\hat H_0$ is the spherical single-particle
part of the  Nilsson potential \cite{Ni69}. The strength parameter $\chi$  in Eq. (4) is fixed using the self-consistent condition \cite{KY95}.
\begin{eqnarray}
\chi_{\tau \tau^{\prime}}= \frac {\frac{2}{3} \varepsilon\hbar \omega_\tau \hbar \omega_{\tau^{\prime}}}{\hbar \omega_n \langle \hat{Q}_0 \rangle_n + \hbar \omega_p \langle \hat{Q}_0 \rangle_p }
\label{xappa}
\end{eqnarray}
The above equation follows from the fact that the quadrupole term in the Nillson model is the mean-field of quadrupole-quadrupole interaction.
The pairing interaction is fitted using the parametrization
\begin{eqnarray}
G_M \;=\; \left[\, G_1 \;\mp\; G_2 \,\frac{N-Z}{A} \,\right] \frac{1}{A}
\label{gm}
\end{eqnarray}
where  ``+ (-)'' is for protons (neutrons), and $ G_{1}$ and  $G_{2}$ are coupling constants which are fitted to the odd-even mass differences.
The parameters are $ G_{1} = 22.68$ and  $G_{2} = 16.22$.
%.   transition 

The electromagnetic transition probabilities from an initial state $\psi^{\sigma_i I_i}$ to a final state $\psi^{\sigma_f I_f}$ are obtained using the expression \cite{SUN19941}$:$
\begin{eqnarray}\label{BE2}
  B(E2, I_i\rightarrow I_f)={e^2 \over {2I_i+1}}\ack \langle \psi^{\sigma_f I_f}\lVert \hat  Q_2\rVert \psi^{\sigma_i I_i}\rangle \ack^2 .
\end{eqnarray}
%from an initial state $\psi^{\sigma_i I_i}$ to a final state $\psi^{\sigma_f I_f}$. %Effective charges of 1.5e (0.5e) for protons (neutrons) similar to our previous publications \cite{GH14,bh15,SJ18} are used in our calculations.
The reduced
magnetic dipole transition probability B(M1) is evaluated using
\begin{eqnarray}\label{BM1}
  B(M1, I_i\rightarrow I_f)={\mu^{2}_{N} \over {2I_i+1}}\ack \langle \psi^{\sigma_f I_f} \lVert  \hat {\mathcal{M}}_1\rVert \psi^{\sigma_i I_i}\rangle \ack^2 ,
\end{eqnarray}
where the magnetic dipole operator is defined as
\begin{eqnarray}\label{M1}
\mathcal{M}^{\tau}_{1}
= g^{\tau}_{l}\, \hat{j}^{\tau}
+ \left( g^{\tau}_{s} - g^{\tau}_{l} \right)\, \hat{s}^{\tau} \, .
\end{eqnarray}
For an irreducible spherical tensor, $\hat  Q$, of rank $L$, the reduced matrix element can be expressed as
\begin{eqnarray}\label{tensor}
  &&\langle \psi^{\sigma_f I_f} \lVert \hat  Q_L \rVert \psi^{\sigma_i I_i}\rangle\nonumber\\
&& ~~~~~~~~=\sum_{\kappa_i,\kappa_f,K_i,K_f}f_{\kappa_iK_i}^{\sigma_i I_i}f_{\kappa_fK_f}^{\sigma_f I_f}\sum_{M_i,M_f,M}(-)^{I_f-M_f}\nonumber\\
&&~~~~~~~~~~~~~~~~~~~~~~~~\times\begin{pmatrix}I_f&L&I_i\\-M_f&M&M_i\end{pmatrix}\nonumber\\
&&~~~~~~~~~~~~~~~~\times\langle \Phi_{\kappa_f}\ack \hat P_{K_fM_f}^{I_f}\hat  Q_{LM}\hat P_{K_iM_i}^{I_i}\ack\Phi_{\kappa_i}\rangle\\
  &&~~~~~~~~ =2\sum_{\kappa_i,\kappa_f,K_i,K_f}f_{\kappa_iK_i}^{\sigma_i I_i}f_{\kappa_fK_f}^{\sigma_f I_f}\sum_{M^\prime,M^{\prime\prime}}(-)^{I_f-K_f}(2I_f+1)^{-1}\nonumber\\
&&~~~~~~~~~~~~~~~~ \times\begin{pmatrix}I_f&L&I_i\\-K_f&M^\prime&M^{\prime\prime}\end{pmatrix}\int d\Omega D_{M^{\prime\prime}K_i}^{I_i}(\Omega)\nonumber\\
&&~~~~~~~~~~~~~~~~~~~~~~~~\times\langle \Phi_{\kappa_f}\ack \hat  Q_{LM^\prime}\hat R(\Omega)\ack\Phi_{\kappa_i}\rangle.\nonumber
\end{eqnarray}
The symbol (   ) in the above expression represents a $3j$-coefficient.
%===============  table 1  ========================
%\begin{table}
%\LTcapwidth=0.4\textwidth
%\caption{ Axial  quadrupole deformation  ($\epsilon$) and Pairing gap parameters ($\Delta_{n}$ and $\Delta_{p}$)
%   employed in the PSM calculation. Axial deformations $\epsilon$ have been considered from the refs. %\cite{Moller1995,moller08,Raman11,101pd,Zhou2011,103pdamr,Cd94,def,def2,107cd15,chiara20,regan1885}. }

%  \begin{tabular}{p{1.1cm}p{1.7cm}p{1.3cm}p{1.3cm}p{1.3cm}p{1.3cm}}%{M{2cm}M{2cm}M{2cm}M{2cm}M{2cm}M{2cm}M{2cm}M{2cm}%M{2cm}M{2cm}N}
%  \hline\hline
%  Isotope	&	$^{101}$Pd	&	$^{103}$Pd	&	$^{107}$Pd	&	$^{109}$Pd	&	$^{111}$Pd\\\hline	
%$\epsilon$	&	0.128 (0.165)	&	0.180	&	0.220	&	0.200	&	0.230	\\	
%$\Delta_{n}$	&0.657 		&	0.957	&	0.684	&	1.025	&	0.827	\\	
%$\Delta_{p}$	&0.772 		&	0.995	&	0.742	&	0.912	&	0.901	\\\hline\hline	
%																						
%Isotope	&	$^{103}$Cd	&	$^{105}$Cd	&	$^{107}$Cd	&	$^{109}$Cd	&	$^{111}$Cd\\	\hline	
%$\epsilon$	&	0.150		&	0.140	&	-0.110	&	0.140	&	0.140\\	
%$\Delta_{n}$	&	0.568	&	0.586	&	1.142	&	1.125	&	0.992	\\	
%$\Delta_{p}$	&	0.756	&	0.897	&	0.878	&	0.947	&	0.806	\\\hline\hline	
												
%\end{tabular}\label{tab1}
%\end{table}
%\vspace{-0.5cm}
%===============  table 1  ========================
\begin{table}
%\LTcapwidth=0.4\textwidth
\caption{ Axial  quadrupole deformation  ($\epsilon$) and pairing gap parameters ($\Delta_{n}$ and $\Delta_{p}$)
   employed in the PSM calculation. Axial deformations $\epsilon$ have been considered from the refs. \cite{Moller1995,moller08,Raman11,101pd,Zhou2011,103pdamr,Cd94,def,def2,107cd15,chiara20,regan1885}. %with some adjustment as discussed in the text. %The nonaxial values ($\gamma$) 
  %are chosen in such a way that observed data is reproduced.
}
%\resizebox{1\columnwidth}{!}
%\large
%\tiny
  \begin{tabular}{p{1.1cm}p{1.7cm}p{1.3cm}p{1.3cm}p{1.3cm}p{1.3cm}}%{M{2cm}M{2cm}M{2cm}M{2cm}M{2cm}M{2cm}M{2cm}M{2cm}M{2cm}M{2cm}N}
  \hline\hline
	&	$^{101}$Pd	&	$^{103}$Pd	&	$^{107}$Pd	&	$^{109}$Pd	&	$^{111}$Pd\\\hline	
$\epsilon$	&	0.128 (0.165)	&	0.180	&	0.220	&	0.200	&	0.230	\\	
$\Delta_{n}$	&0.657 		&	0.957	&	0.684	&	1.025	&	0.827	\\	
$\Delta_{p}$	&0.772 		&	0.995	&	0.742	&	0.912	&	0.901	\\\hline\hline

	&	$^{103}$Cd	&	$^{105}$Cd	&	$^{107}$Cd	&	$^{109}$Cd	&	$^{111}$Cd\\	\hline	
$\epsilon$	&	0.150		&	0.140	&	-0.110	&	0.140	&	0.140\\	
$\Delta_{n}$	&	0.568	&	0.586	&	1.142	&	1.125	&	0.992	\\	
$\Delta_{p}$	&	0.756	&	0.897	&	0.878	&	0.947	&	0.806	\\\hline\hline	
												
%Isotope  & $^{101}$Pd & $^{103}$Pd  & $^{105}$Pd  & $^{107}$Pd  & $^{109}$Pd & $^{111}$Pd &$^{101}$Cd & $^{103}$Cd  & $^{105}$Cd  & $^{107}$Cd  & $^{109}$Cd & $^{111}$Cd\\
%\hline $\epsilon$ & 0.165     & 0.180       & 0.170      &  0.220     &  0.200    & 0.230 & 0.18& 0.15 & 0.14 & -0.11 & 0.14 & 0.14\\
%$\Delta_{n}$    & 0.526 & 0.957 & 1.154 & 0.684 & 1.025 & 0.827 & 0.637 & 0.568 & 0.586 &1.142 &1.125 & 0.992 \\
%$\Delta_{p}$    & 0.738 & 0.995 & 0.943 & 0.742 & 0.912 & 0.901 & 0.971 & 0.756 & 0.897 & 0.878 & 0.947 & 0.806 \\\hline\hline
%% Isotope  &$^{101}$Cd & $^{103}$Cd  & $^{105}$Cd  & $^{107}$Cd  & $^{109}$Cd & $^{111}$Cd\\
%% \hline $\epsilon$ & 0.165     & 0.180       & 0.170      &  0.220     &  0.200    &0.230 \\
%% $\Delta_{n}$    &0.526 &0.957& 1.154 & 0.684 & 1.025 & 0.827 \\
%%   $\Delta_{p}$    &0.738 &0.995& 0.943 & 0.742 & 0.912 & 0.901 \\\hline\hline
%\end{tabular}\label{tab1}
%\end{table}
\end{tabular}\label{tab1}
\end{table}
%====================================================
\section{Semiclassical particle-rotor model}
\label{scm}
\begin{figure}[!b]
\centering
\includegraphics[width=0.9\columnwidth]{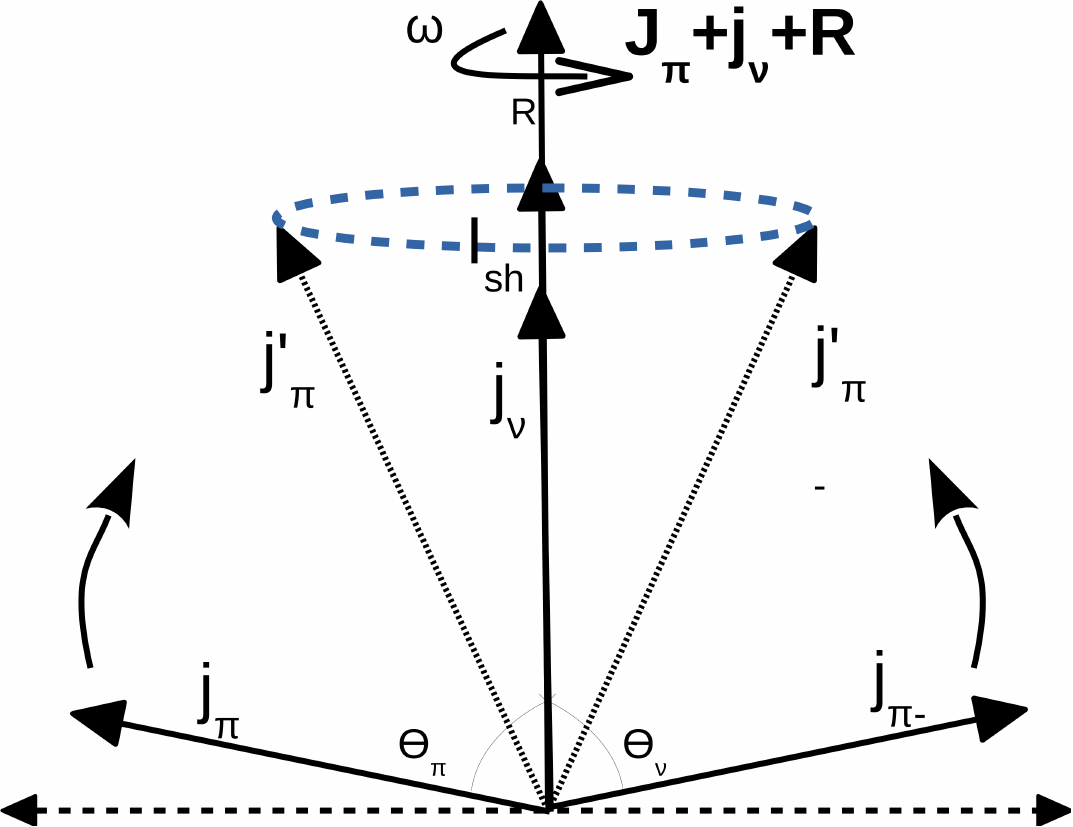}
\caption{Schematic illustration of the twin-shears mechanism leading to the formation of an antimagnetic rotational band.}
\label{AMR}
\end{figure}

The semiclassical model (SCM), originally developed by Macchiavelli \emph{et al.}~\cite{Macchiavelli,macch1} to describe magnetic rotation, was later modified by Sugawara \emph{et al.}~\cite{144dy} to explain antimagnetic rotational bands. In this framework, the angular momentum of the neutron particles is denoted by $\vec{j}_{\nu}$, while those of the two proton holes are $\vec{j}_{\pi}^{\,1}$ and $\vec{j}_{\pi}^{\,2}$. The angle between $\vec{j}_{\nu}$ and each proton-hole angular momentum vector is the shears angle $\theta$. The total angular momentum is generated through the gradual closing of the two shears blades in a back-to-back symmetric fashion, as illustrated in Fig.~\ref{AMR}. Owing to this symmetry, antimagnetic rotational bands consist of levels differing in spin by $2\hbar$.

The total energy $E(I(\theta))$ is expressed as the sum of the rotational energy of the core and the effective interaction energy between the shears blades,
%\begin{equation}
\begin{align}
E(I(\theta)) = &\frac{R^{2}}{2\Im^{(2)}} + V_{2}P_{2}(\theta) \nonumber  \\
=& \frac{(I - j_{\pi} - j_{\nu})^{2}}{2\Im^{(2)}} + V_{2}P_{2}(\theta).
%\end{equation}
\end{align}
Expanding the interaction term to include proton--neutron and proton--proton interactions, the total energy can be written as
\begin{align}
E(I) =
&\frac{(I - j_{\pi} - j_{\nu})^{2}}{2\Im^{(2)}}
+ \frac{V_{\pi\nu}}{2}(3\cos^{2}\theta - 1) \nonumber  \\
&- \frac{V_{\pi\pi}}{n}\,\frac{(3\cos^{2}(2\theta) - 3)}{2}.
\end{align}

Here, $\Im^{(2)}$ denotes the dynamic moment of inertia of the core, $V_{\pi\nu}$ and $V_{\pi\pi}$ represent the effective proton--neutron and proton--proton interaction strengths, respectively, and $n$ is the number of particle--hole pair combinations for a given configuration~\cite{110Cd}. The first term corresponds to core rotation, the second term arises from the repulsive interaction between proton holes and neutron particles, and the third term represents the attractive interaction between proton holes.

The total angular momentum is obtained by minimizing the energy with respect to the shears angle,
\begin{equation}
\frac{dE(I)}{d\theta} = 0,
\end{equation}
which leads to
\begin{align}
I = & a j + 2j\cos\theta
+ \frac{1.5\,\Im^{(2)} V_{\pi\nu}\cos\theta}{j} \nonumber  \\
&- \frac{6\,\Im^{(2)} V_{\pi\pi}\cos2\theta\cos\theta}{n j},
\end{align}
where $a = j_{\nu}/j_{\pi}$ and $j = j_{\pi}$. The first two terms, $aj + 2j\cos\theta$, represent the contribution from the pure shears mechanism.

At the bandhead ($\theta = 90^{\circ}$), the angular momentum reduces to $I = aj = j_{\nu}$, indicating the alignment of the neutron angular momentum. As the shears gradually close with increasing spin, the angular momentum builds up and reaches the maximum shears angular momentum $I_{sh}^{\mathrm{max}}$ at full closure ($\theta = 0^{\circ}$).

The angular frequency associated with the shears mechanism, $\omega_{sh}$, is given by
\begin{align}
\omega_{sh}
=& \left( \frac{dE_{sh}}{d\theta} \right)
\Big/ \left( \frac{dI_{sh}}{d\theta} \right)  \nonumber  \\
= &\frac{1.5 V_{\pi\nu}\cos\theta}{j}
- \frac{6 V_{\pi\pi}\cos2\theta\cos\theta}{n j}.
\end{align}

The total angular momentum can thus be written as
\begin{equation}
I = I_{sh} + \Im^{(2)} \omega_{sh}.
\end{equation}

At full shears closure ($\theta = 0^{\circ}$),
\begin{equation}
\Im^{(2)} \omega_{sh}\big|_{\theta = 0}
= I^{\mathrm{max}} - I_{sh}^{\mathrm{max}},
\end{equation}
which gives
\begin{equation}
\omega_{sh}\big|_{\theta = 0}
= \frac{1.5 V_{\pi\nu}}{j}
- \frac{6 V_{\pi\pi}}{n j}.
\end{equation}

The total rotational frequency is expressed as
\begin{equation}
\omega_{rot} = \omega + \omega_{sh}.
\end{equation}

The reduced transition probability for the electric quadrupole transition is given as a function of the shears angle by
\begin{equation}
B(E2) = \frac{15}{32\pi} (eQ_{\mathrm{eff}})^{2} \sin^{4}\theta,
\end{equation}
where the effective quadrupole moment is
\begin{equation}
eQ_{\mathrm{eff}}
= e_{\pi} Q_{\pi}
+ \left( \frac{j_{\pi}}{j_{\nu}} \right)^{2}
e_{\nu} Q_{\nu}.
\end{equation}
In the $A \approx 100$ mass region, the effective interaction strengths are typically
$V_{\pi\nu} \approx 1.2$~MeV and $V_{\pi\pi} \approx 0.2$~MeV.

%% %================================leveldiagrm========================
%===========fig16=====================1===================
\begin{figure}[!t]
 \centerline{\includegraphics[trim=0cm 0cm 0cm
0cm,width=0.50\textwidth,clip]{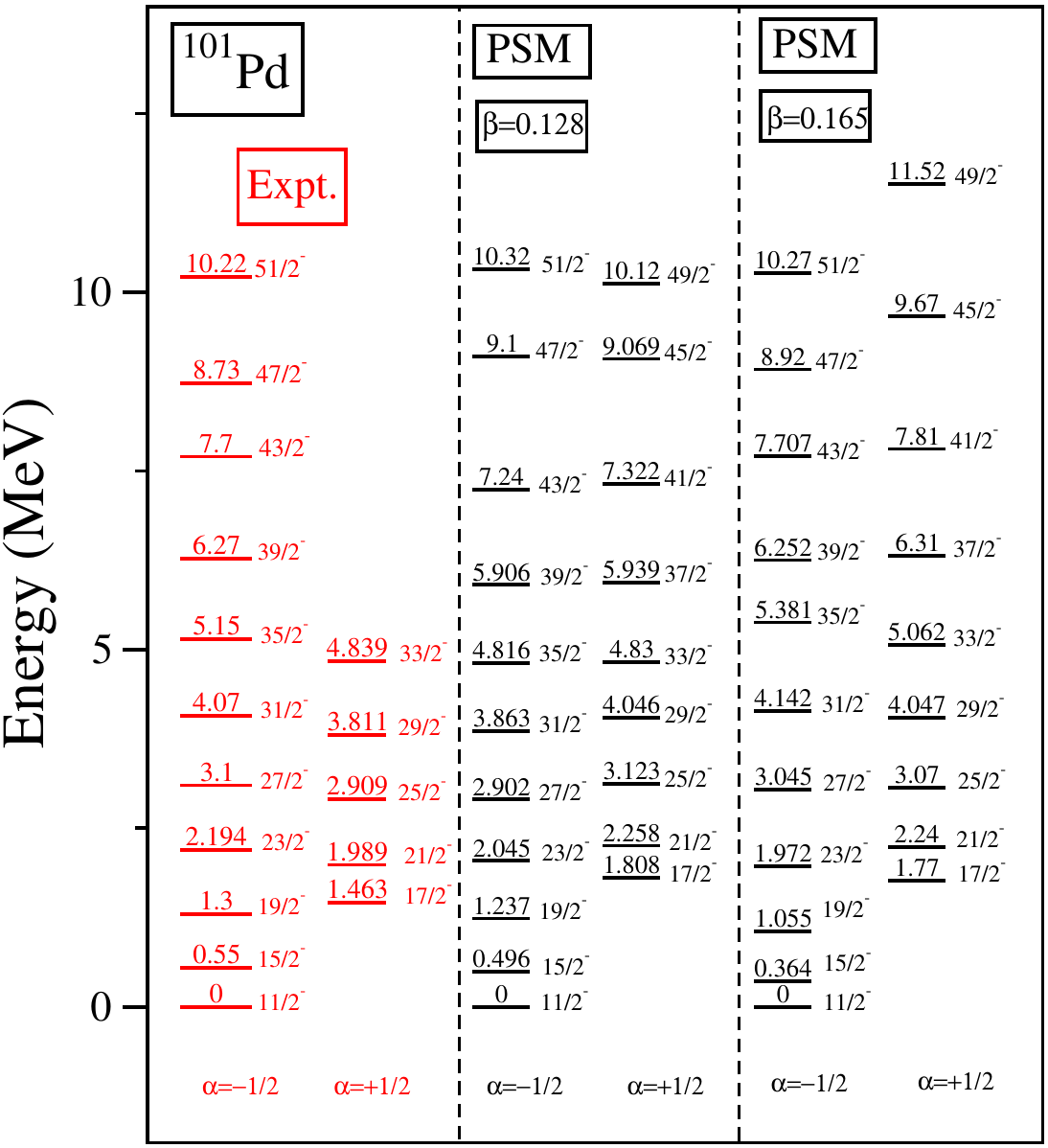}} \caption{(Color
   online) Comparison of PSM calculated energies after configuration mixing with the corresponding available experimental data \cite{101pdsuga2} for $^{101}$Pd.
 }
\label{levelpd1}
\end{figure}
%========================================================
%===========fig16=====================1===================
\begin{figure}[htb]
 \centerline{\includegraphics[trim=0cm 0cm 0cm
0cm,width=0.50\textwidth,clip]{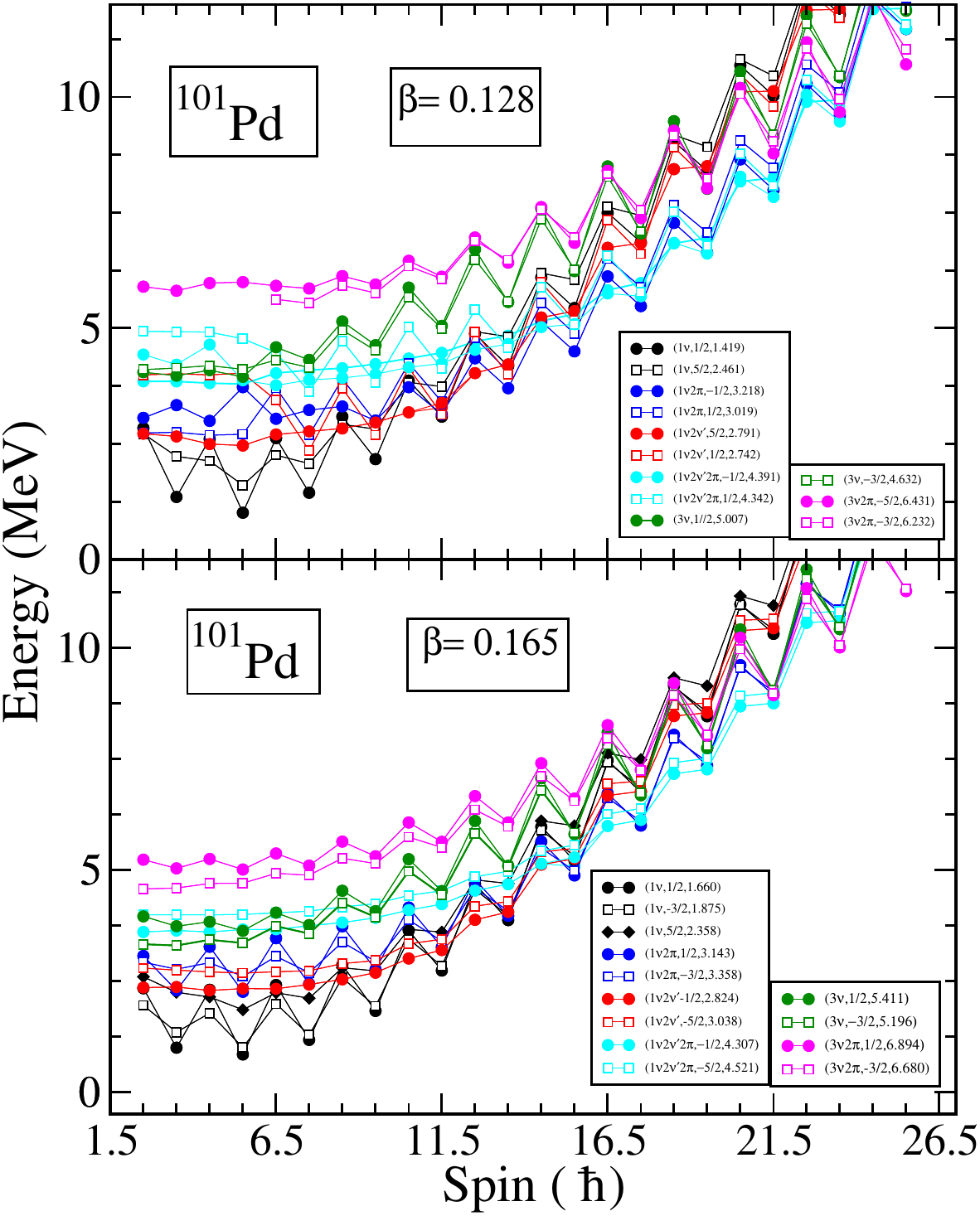}} \caption{(Color
   online) Angular-momentum projected energies are shown before diagonalization of the shell model Hamiltonian for $^{101}$Pd.
 }
\label{bd_pd1}
\end{figure}
%========================================================
%===========fig16=====================1===================
\begin{figure*}[htb]
 \centerline{\includegraphics[trim=0cm 0cm 0cm
0cm,width=16cm,height=10cm,clip]{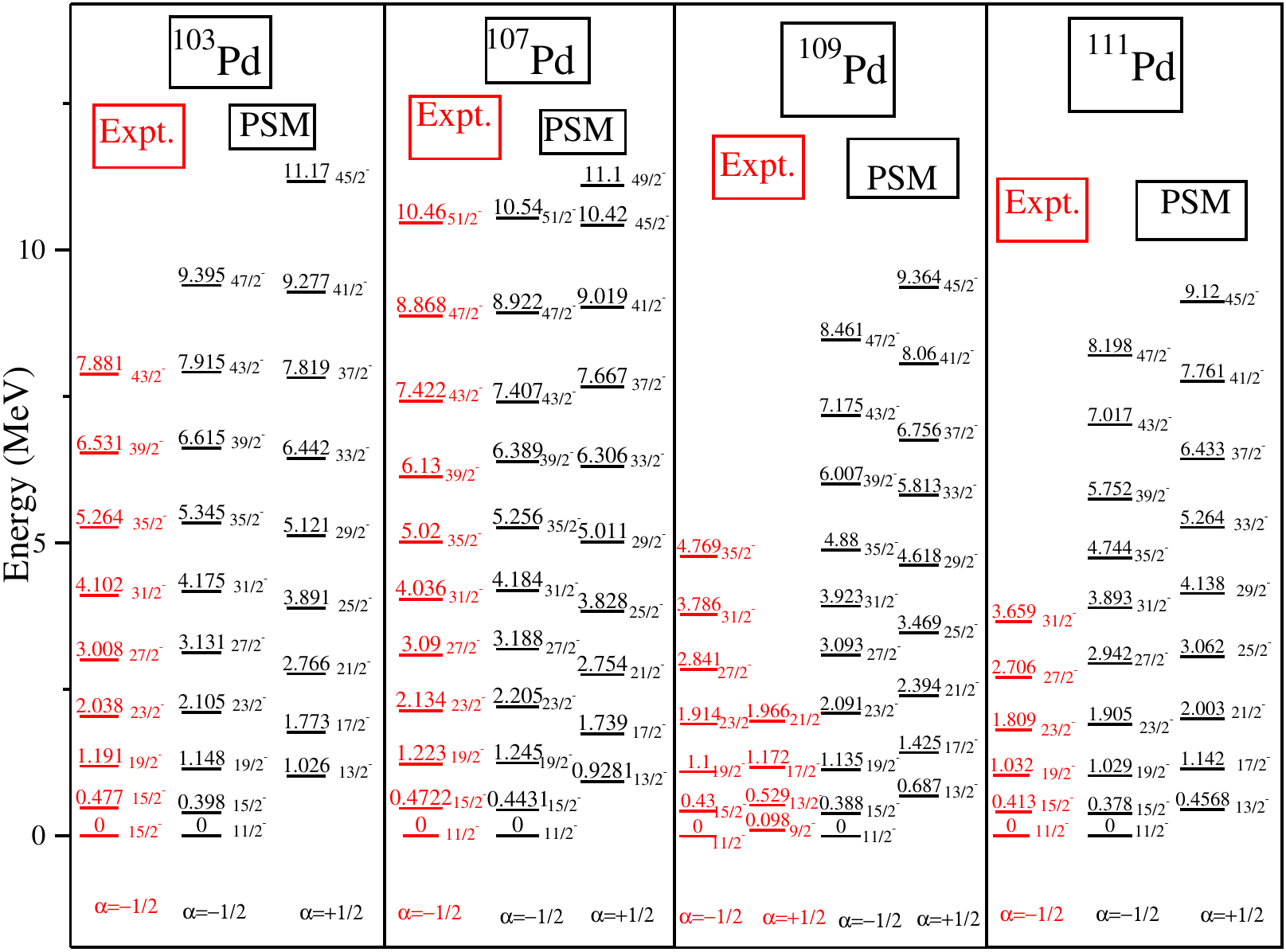}} \caption{(Color
   online) Comparison of PSM calculated energies after configuration mixing with the corresponding available experimental data \cite{103pdamr,Cd94,109pd} for $^{103,107-111}$Pd.
 }
\label{levelpd2}
\end{figure*}
%===========fig16=====================1===================
%===========fig17=====================1===================
\begin{figure}[htb]
 \centerline{\includegraphics[trim=0cm 0cm 0cm
0cm,width=0.50\textwidth,height=16cm,clip]{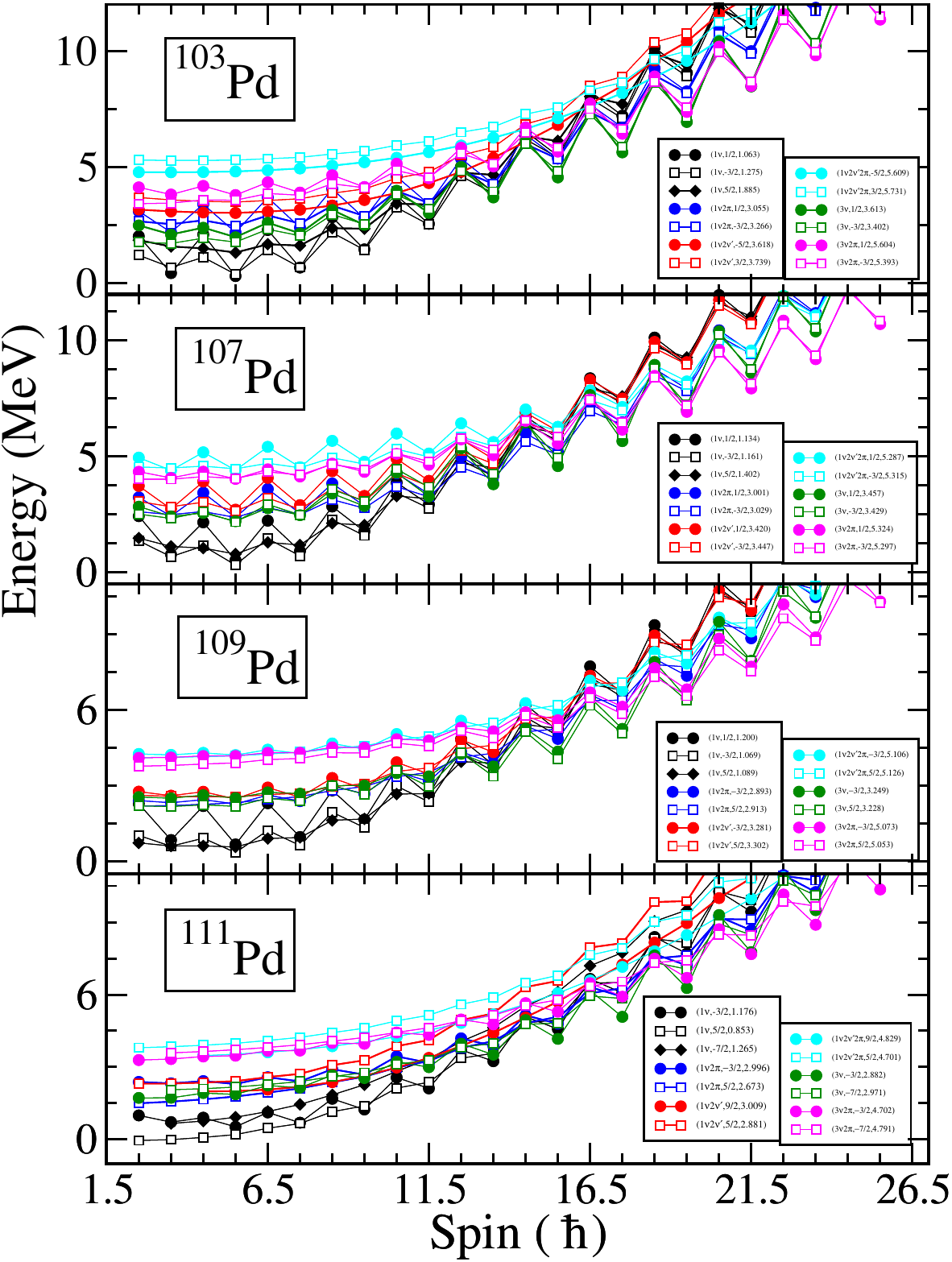}} \caption{(Color
   online) Angular-momentum projected energies are shown before diagonalization of the shell model Hamiltonian for $^{103,107-111}$Pd.
 }
\label{bd_pd2}
\end{figure}
%========================================================
%===========fig16=====================1===================
%\begin{figure}[htb]
% \centerline{\includegraphics[trim=0cm 0cm 0cm
%0cm,width=0.40\textwidth,height=10cm,clip]{107pd_exthe.eps}} \caption{(Color
 %  online) Comparison of PSM calculated energies after configuration mixing with the corresponding available experimental data \cite{109pd} for $^{109-111}$Pd.
% }
%\label{levelpd2n}
%\end{figure}
%===========fig16=====================1===================

\begin{figure}[htb]
 \centerline{\includegraphics[trim=0cm 0cm 0cm
0cm,width=0.50\textwidth,clip]{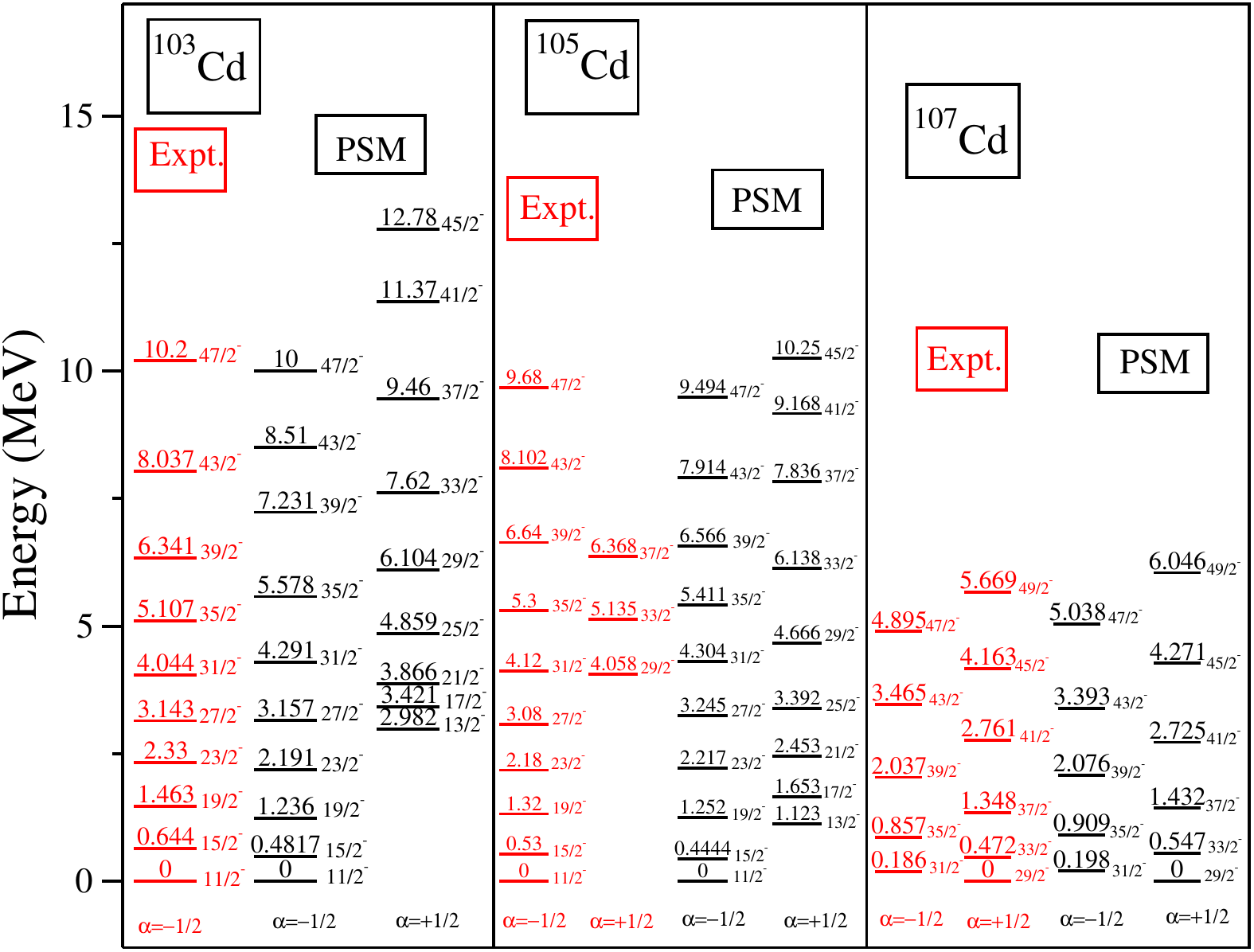}} \caption{(Color
   online) Comparison of PSM calculated energies after configuration mixing with the corresponding available experimental data \cite{103cdbt,105cdamr,107cd15} for $^{103-107}$Cd.
 }
\label{levelcd1}
\end{figure}
%========================================================
%===========fig19=====================1===================
\begin{figure}[htb]
 \centerline{\includegraphics[trim=0cm 0cm 0cm
0cm,width=0.50\textwidth,height=16cm,clip]{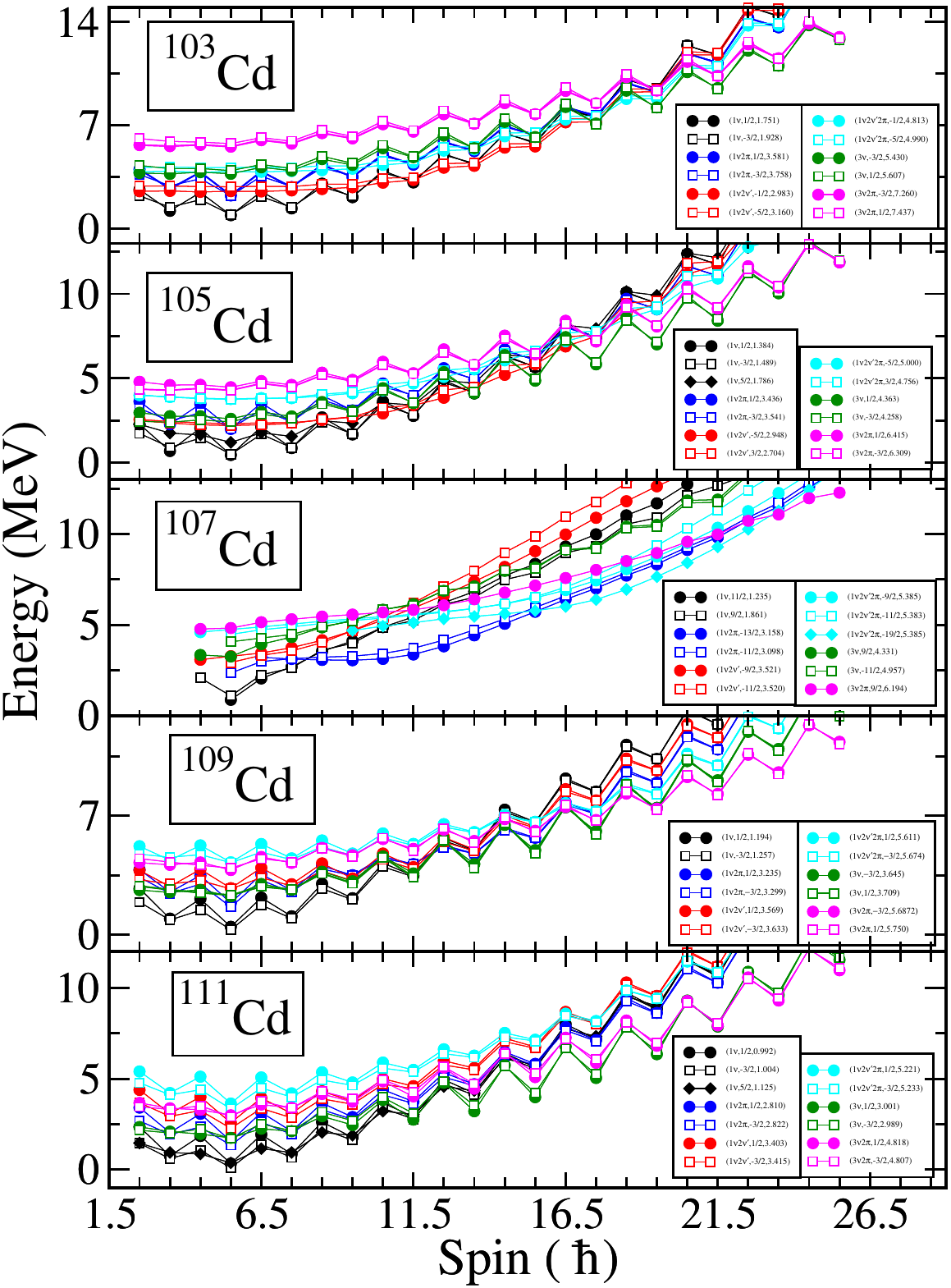}} \caption{(Color
   online) Angular-momentum projected energies are shown before diagonalization of the shell model Hamiltonian for $^{103-111}$Cd.
 }
\label{bd_cd2}
\end{figure}
%========================================================

%===========fig16=====================1===================
\begin{figure}[htb]
 \centerline{\includegraphics[trim=0cm 0cm 0cm
0cm,width=0.45\textwidth,clip]{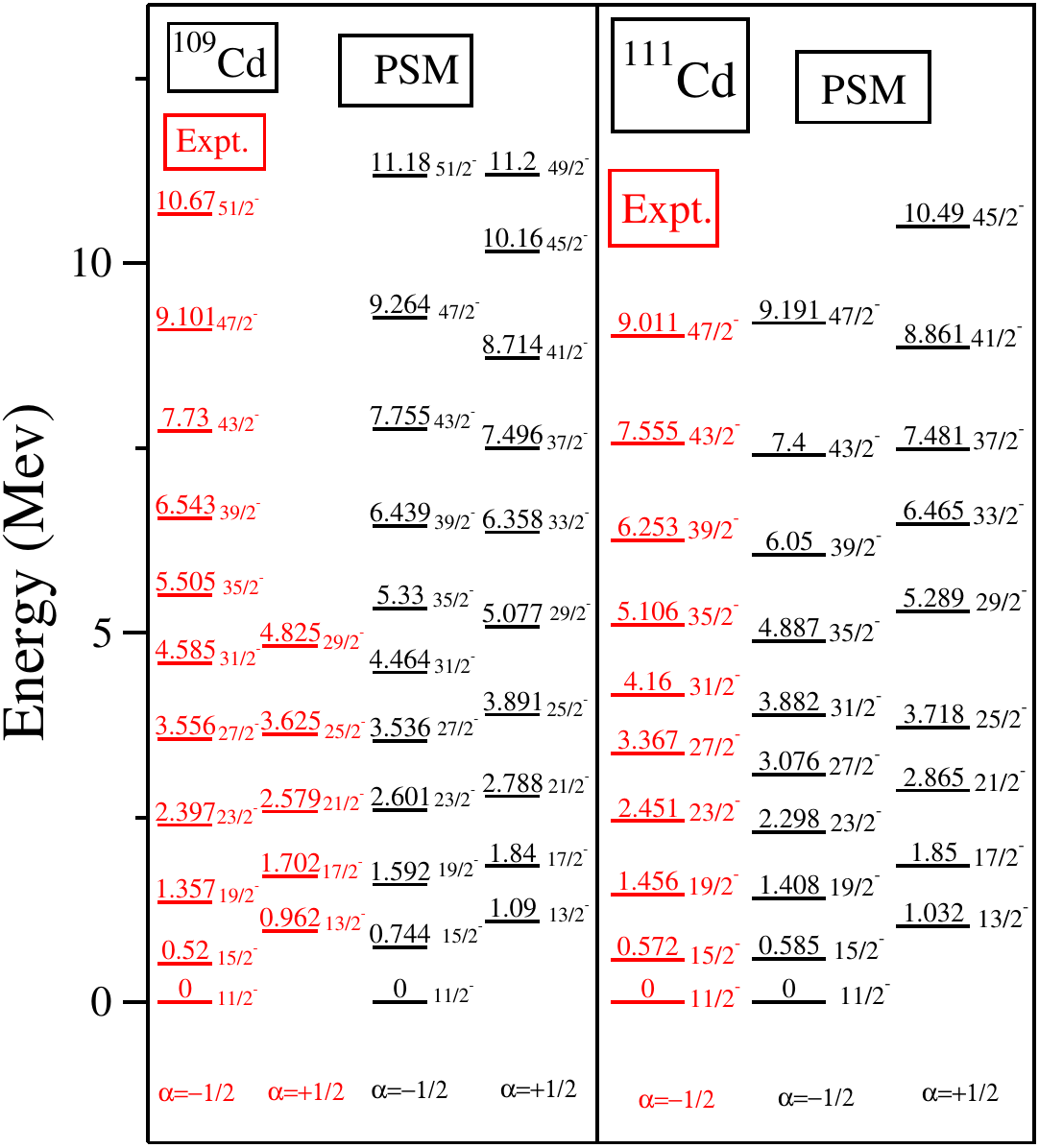}} \caption{(Color
   online) Comparison of PSM calculated energies after configuration mixing with the corresponding available experimental data \cite{chiara20,109cd} for $^{109-111}$Cd.
 }
\label{levelcd2}
\end{figure}

%%%%%%%%%%%%%%%%%%%%%%%%%%%%%%%%%%%%%%%%%%%%%%%%%%%%%%%%%

%% %===========fig18=====================1===================
%% \begin{figure}[htb]
%%  \centerline{\includegraphics[trim=0cm 0cm 0cm
%% 0cm,width=0.50\textwidth,clip]{109pd_bd.eps}} \caption{(Color
%%    online) Angular-momentum projected energies are shown before diagonalization of the shell model Hamiltonian for $^{109-111}$Pd.
%%  }
%% \label{fig18}
%% \end{figure}
%% %========================================================
%===========fig19=====================1===================
%\begin{figure}[htb]
 %\centerline{\includegraphics[trim=0cm 0cm 0cm
%0cm,width=0.50\textwidth,clip]{101cd_bd.eps}} \caption{(Color
%   online) Angular-momentum projected energies are shown before diagonalization of the shell model Hamiltonian for $^{103-105}$Cd.
% }
%\label{bd_cd1}
%\end{figure}
%========================================================
%% %% ===========fig19=====================1===================
%% \begin{figure}[htb]
%%  \centerline{\includegraphics[trim=0cm 0cm 0cm
%% 0cm,width=0.50\textwidth,clip]{105cd_bd.eps}} \caption{(Color
%%    online) Angular-momentum projected energies are shown before diagonalization of the shell model Hamiltonian for $^{105-107}$Cd.
%%  }
%% \label{fig19}
%% \end{figure}
%% %========================================================

%===========fig4=====================1===================
\begin{figure}[htb]
 \centerline{\includegraphics[trim=0cm 0cm 0cm
0cm,width=0.50\textwidth,clip]{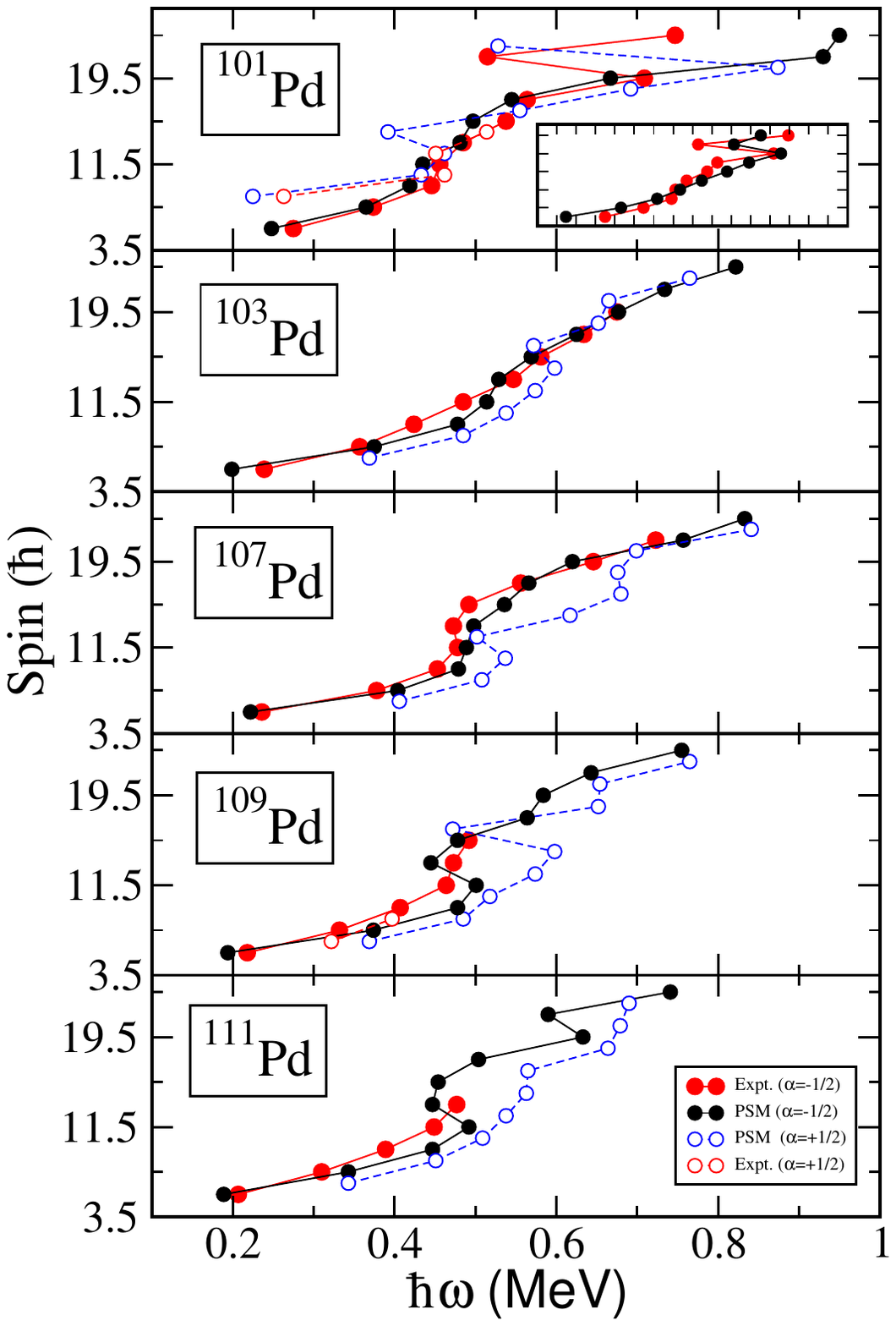}} \caption{(Color
   online) The PSM values of spin ($\hbar$) versus  rotational frequency ($\hbar \omega=\frac{E(I)-E(I-2)}{2}$) are compared with  experimental values for $^{101,103,107-111}$Pd.
 }
\label{ixpd}
\end{figure}
%=========================================================

%===========fig3=====================1===================
\begin{figure}[htb]
 \centerline{\includegraphics[trim=0cm 0cm 0cm
0cm,width=0.50\textwidth,clip]{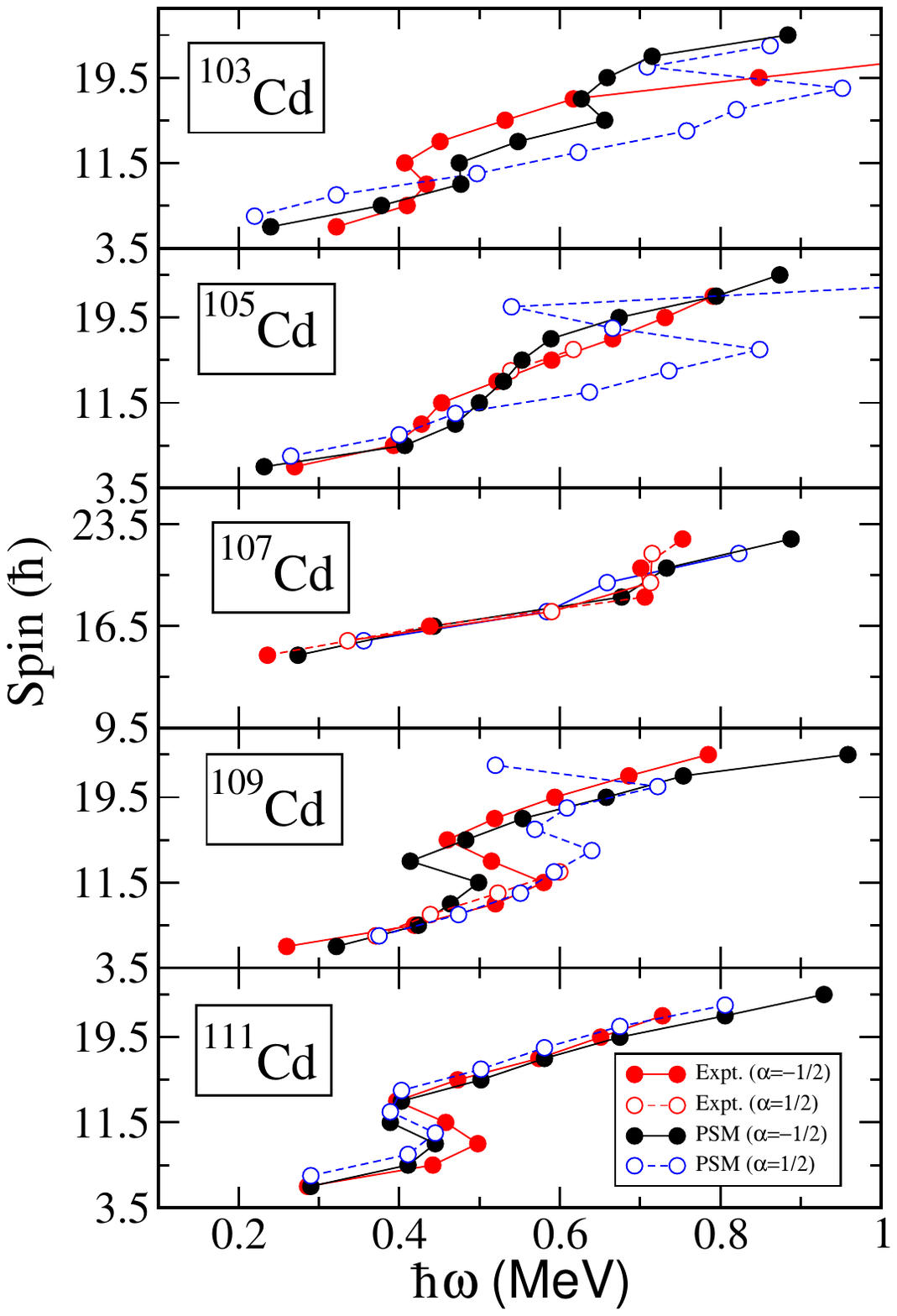}} \caption{(Color
   online) The PSM values of spin ($\hbar$) versus  rotational frequency ($\hbar \omega=\frac{E(I)-E(I-2)}{2}$)  are compared with  experimental values for $^{103-111}$Cd.
 }
\label{ixcd}
\end{figure}
%=========================================================
%.  wavefunction
%===========fig13=====================1===================
 \begin{figure}[htb] 
   \centerline {\includegraphics[trim=0cm 0cm 0cm 0cm,width=0.5\textwidth,clip]{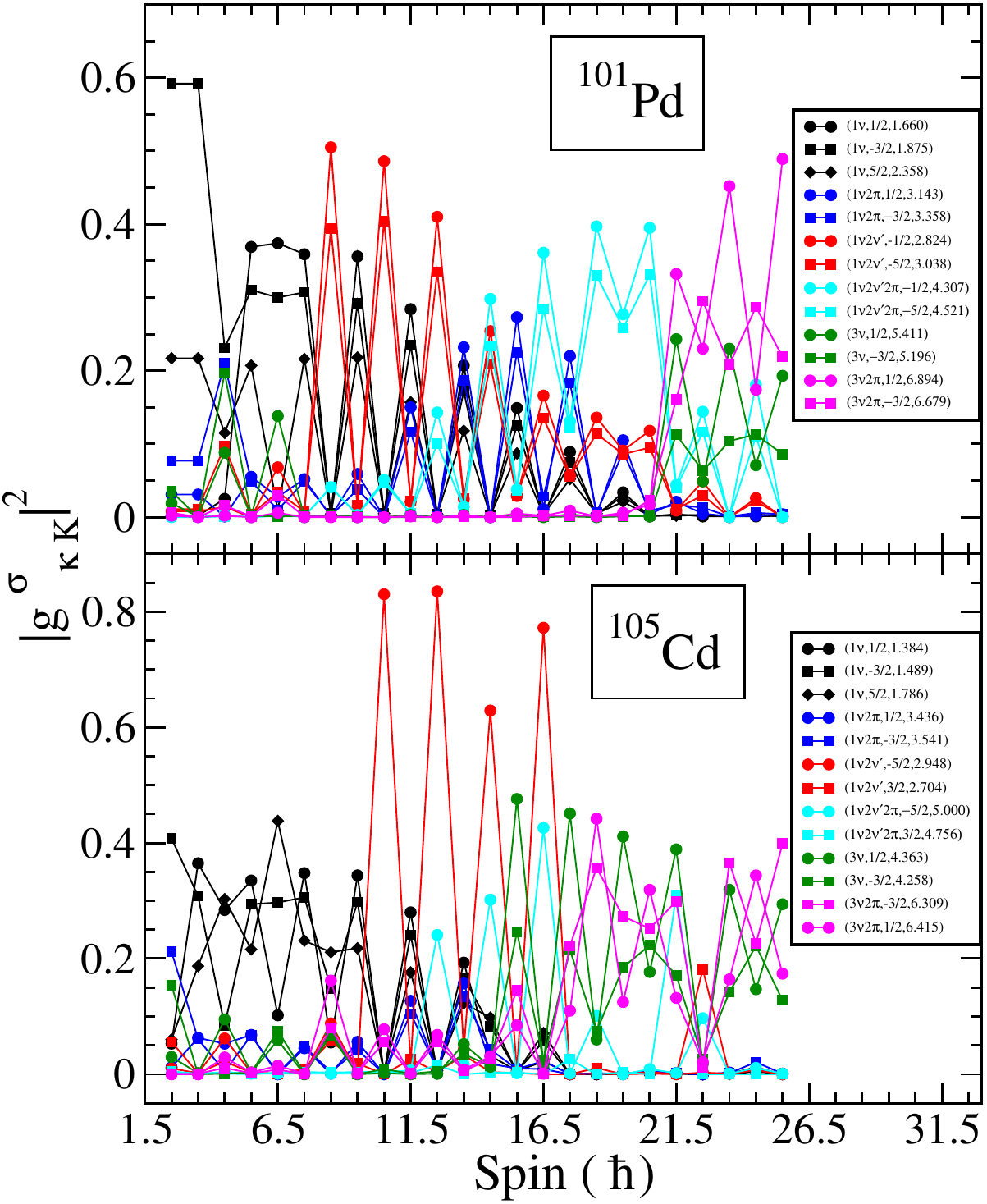}} 
  \caption{(Color
    online) Wavefunction amplitudes of various projected K-configurations
 after diagonalization are plotted for $^{101}$Pd and $^{105}$Cd nuclides.
  }
 \label{wf}
\end{figure}
%% %========================================================
%-----------------figures--------------------
%===========fig12=====================1===================
\begin{figure}[htb]
 \centerline{\includegraphics[trim=0cm 0cm 0cm
0cm,width=0.48\textwidth,clip]{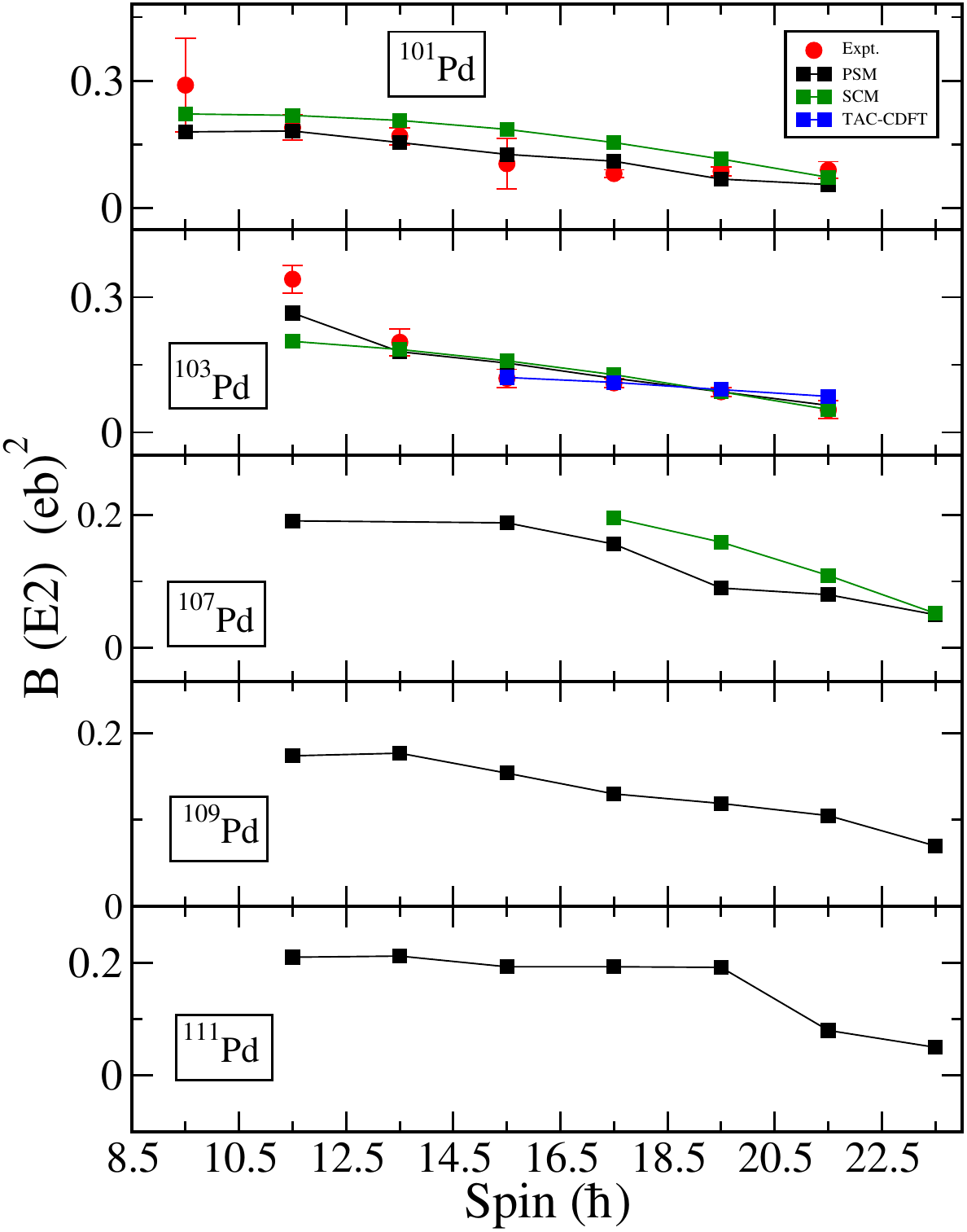}} \caption{(Color online)  Comparison of B(E2) transition probabilities (e$^2$b$^2$) for $^{101,103,107-111}$Pd isotopes.
 }
\label{be2pd}
\end{figure}
%========================================================
%========================================================
%===========fig11=====================1===================
\begin{figure}[htb]
 \centerline{\includegraphics[trim=0cm 0cm 0cm
0cm,width=0.48\textwidth,clip]{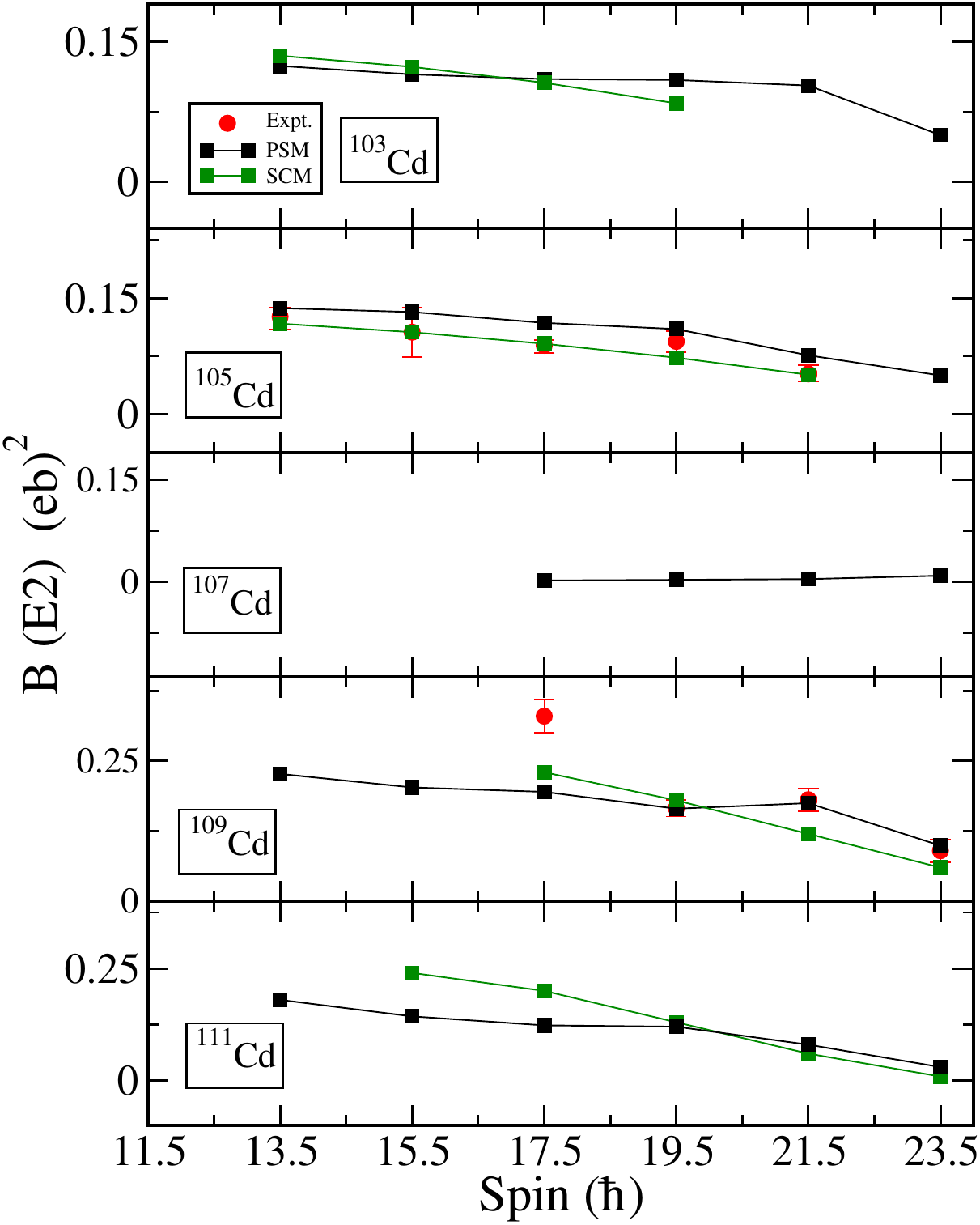}} \caption{(Color online) Comparison of B(E2) transition probabilities (e$^2$b$^2$) for $^{103-111}$Cd isotopes.
 }
\label{be2cd}
\end{figure}
\begin{figure}[htb]
 \centerline{\includegraphics[trim=0cm 0cm 0cm
0cm,width=0.48\textwidth,clip]{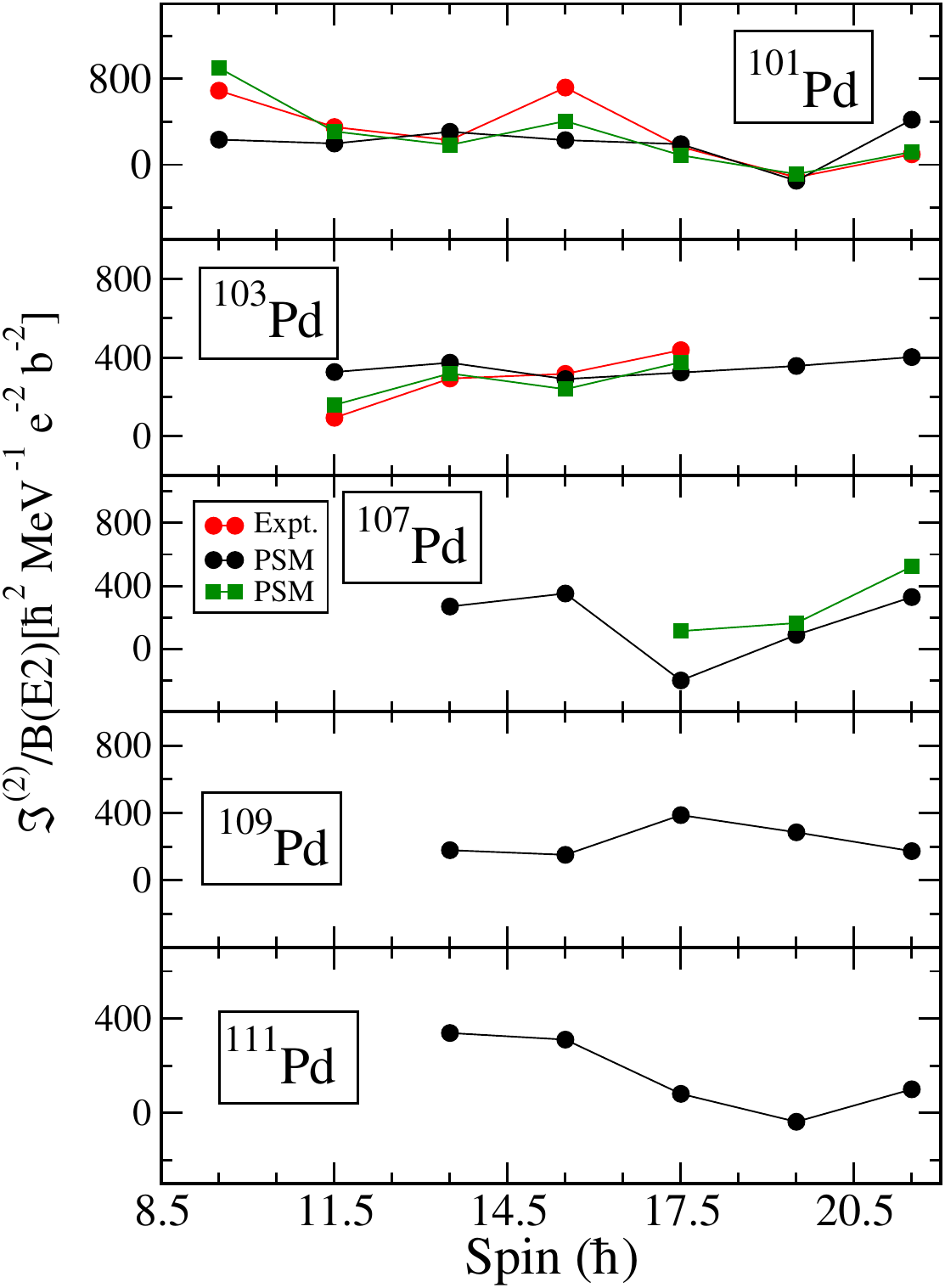}} \caption{(Color
   online) Comparison between experimental and calculated ratio of dynamic moment of inertia, $\mathcal{J}^{(2)}=\frac{4}{E_{\gamma}(I)-E_{\gamma}(I-2)}$, to B(E2) for $^{101,103,107-111}$Pd isotopes.
 }
\label{j2pd}
\end{figure}
%========================================================
%=========================================================
%===========fig9=====================1===================
\begin{figure}[htb]
 \centerline{\includegraphics[trim=0cm 0cm 0cm
0cm,width=0.48\textwidth,clip]{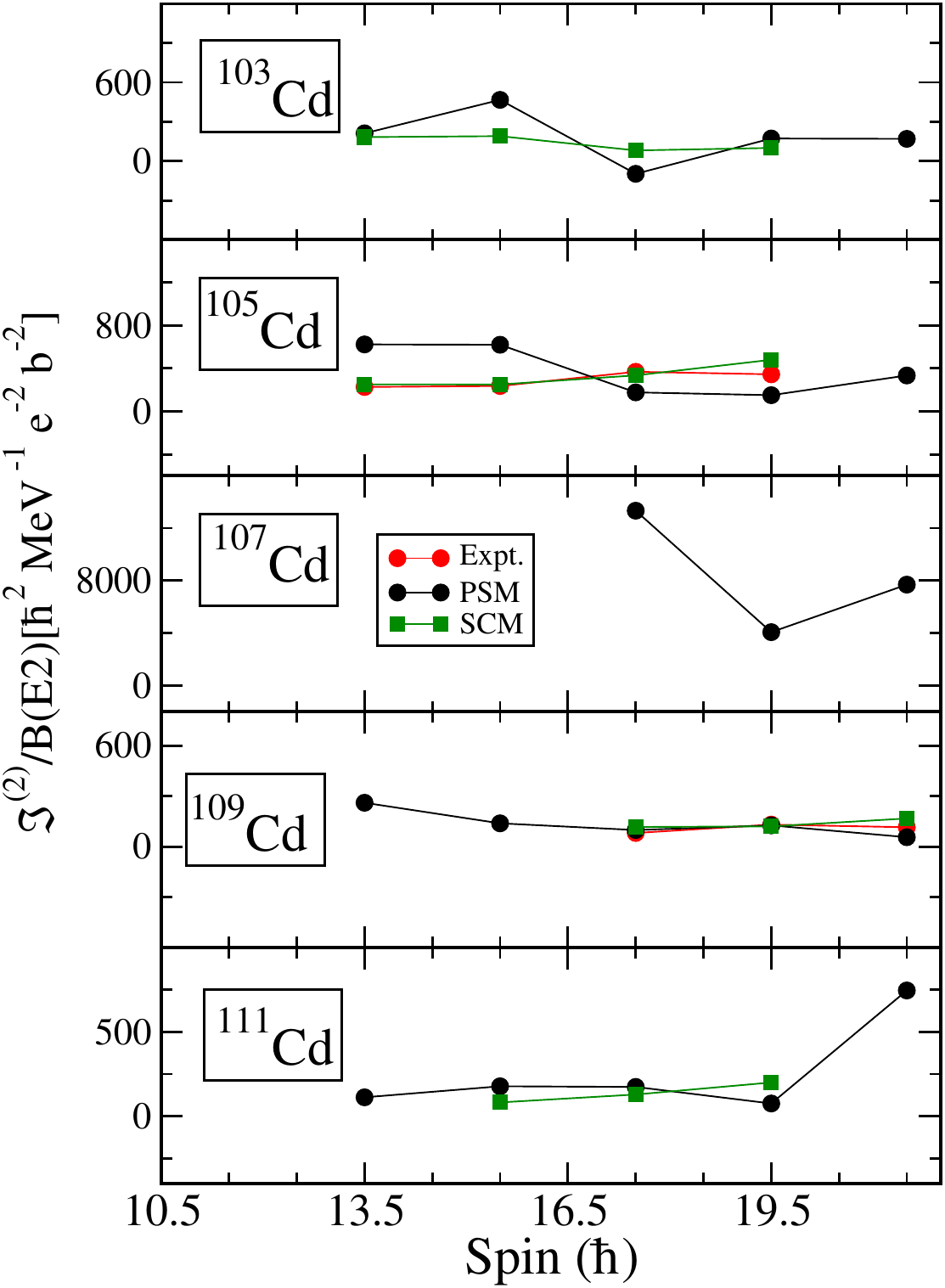}} \caption{(Color
   online) Comparison between experimental and calculated ratio of dynamic moment of inertia, $\mathcal{J}^{(2)}=\frac{4}{E_{\gamma}(I)-E_{\gamma}(I-2)}$, to B(E2) for $^{103-111}$Cd isotopes.
 }
\label{j2cd}
\end{figure}
%% %===========fig2=====================1===================
%% \begin{figure}[htb]
%%  \centerline{\includegraphics[trim=0cm 0cm 0cm
%% 0cm,width=0.45\textwidth,clip]{ENERGY_AMR_PD.pdf}} \caption{(Color
%%    online) Comparison of PSM calculated energies after configuration mixing with the corresponding available experimental data \cite{101pdsuga2,103pdamr,105pd,Cd94,109pd} for $^{101-111}$Pd.
%%  }
%% \label{fig2}
%% \end{figure}
%% %=========================================================
\begin{figure}[htb]
 \centerline{\includegraphics[trim=0cm 0cm 0cm
0cm,width=0.50\textwidth,clip]{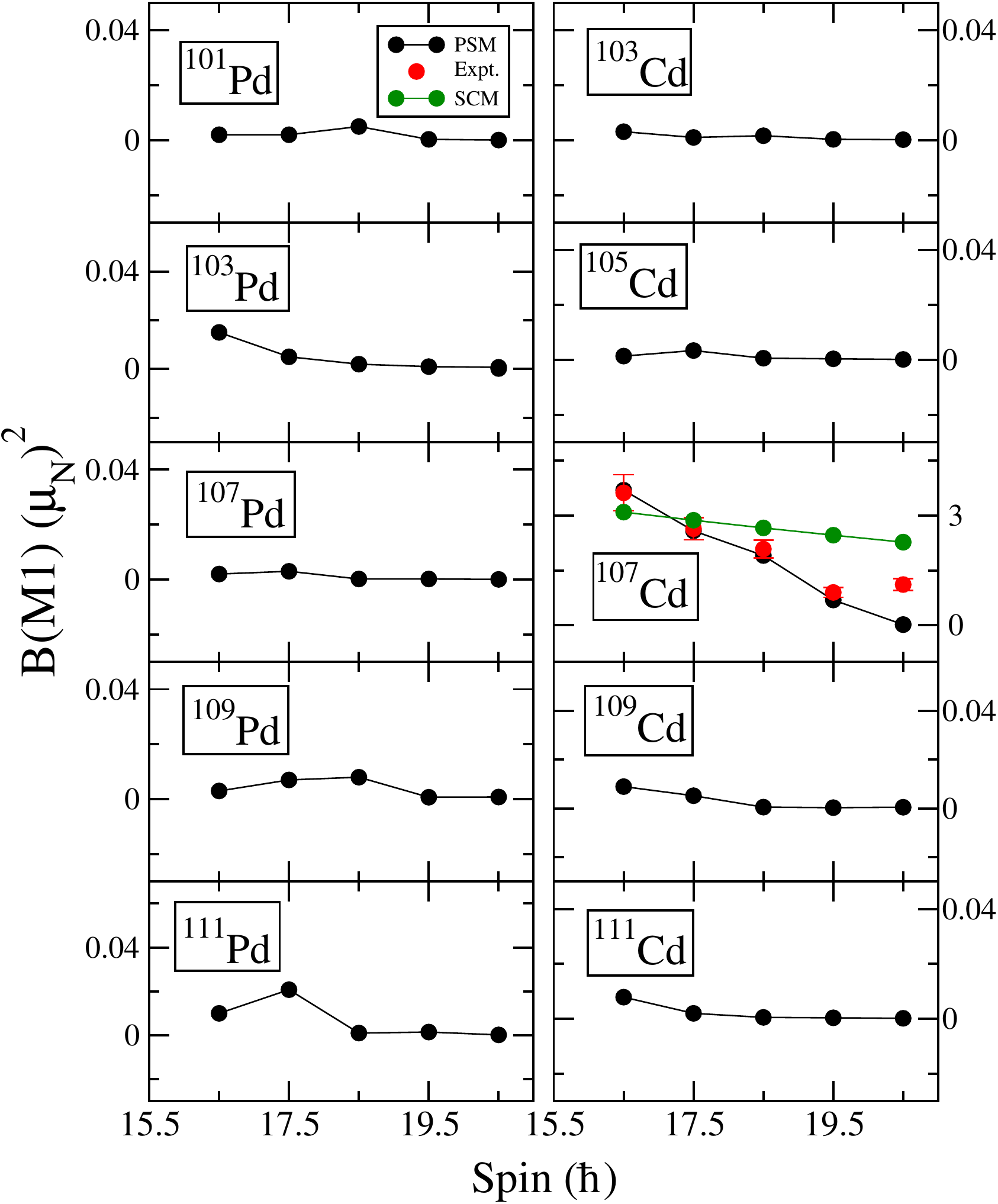}} \caption{(Color online) Comparison of B(M1) transition probabilities %($\mu_{N})^{2}$
 for $^{101,103,107-111}$Pd and  $^{103-111}$Cd isotopes.
 }
\label{bm1}
\end{figure}
\begin{figure}[htb]
 \centerline{\includegraphics[trim=0cm 0cm 0cm
0cm,width=0.43\textwidth,clip]{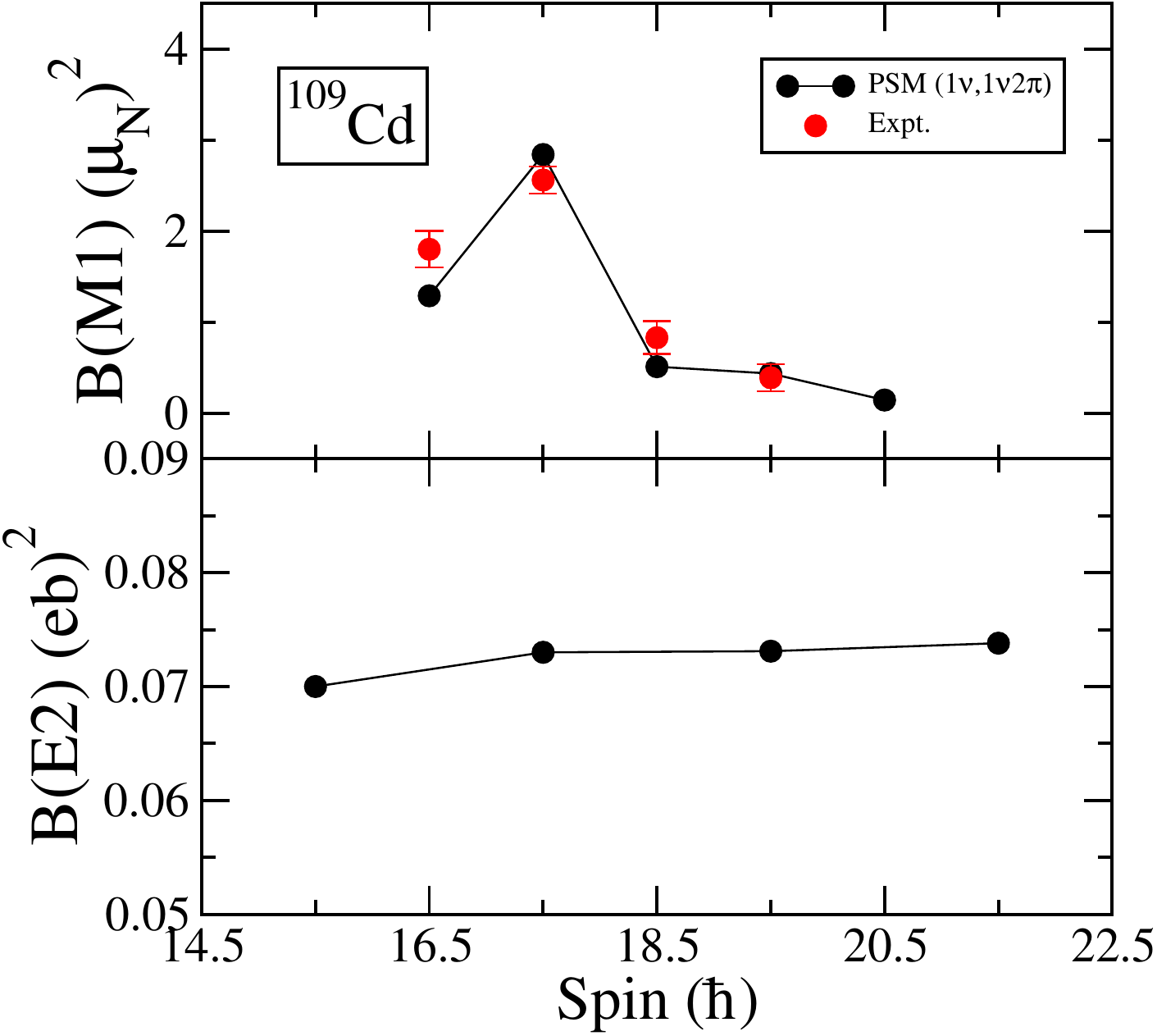}} \caption{(Color online) Comparison of B(M1)  and  B(E2) transition probabilities %($\mu_{N})^{2}$
 for  $^{109}$Cd isotope.
 }
\label{bebm}
\end{figure}
\section{Results and discussion}\label{result}

The extended PSM calculations have been performed for ten nuclides of $^{101,103,107,109,111}$Pd and $^{103,105,107,109,111}$Cd.
We have chosen these nuclei as some of the them have well established MR and AMR band structures and for others predictions are
made for the possible existence of these modes.
It may be noted that
$^{105}$Pd is missing from the studied Pladium series as for this system, wobbling bands have been identified and has
been studied using the triaxial projected shell model approach \cite{Karmakar2024}.
The deformation values used to obtain the Nilsson configurations are listed in Table \ref{tab1}. These
values have been adopted from the earlier investigations on the studied nuclei \cite{Moller1995,moller08,Raman11,101pd,Zhou2011,103pdamr,Cd94,def,def2,107cd15,chiara20,regan1885}. Table \ref{tab1} also
provides the pairing gaps obtained by solving the BCS equations with the strength parameters of the
pairing interaction given in Eq.~(\ref{gm}). In the following subsections, we discuss the results obtained for excitation energy, aligned
angular-momentum and transition probabilities.

\subsection{ Excitation Energy }

In this subsection, a detailed comparison of the PSM calculated excitation energies with the available data shall be made, and the
results will be discussed individually for each nucleus.

The high-spin states of $^{101}$Pd were studied using the in-beam $\gamma$-ray spectroscopic method 
\cite{Zhou2011}. The $\alpha=-1/2$ signature branch of the negative parity band based
on $1/2^- [550]$ Nilsson configuration was extended up to spin, I=$43/2^-$. The $\alpha=+1/2$ branch of the
band was also observed in this work up to I=$33/2^-$. The cranked shell model (CSM) analysis was performed
and it was shown that the first observed bandcrossing at $\hbar \omega = 0.45$MeV can be associated with the
alignment of two-protons in $1g_{9/2}$. In a subsequent work \cite{101pdsuga}, the
negative parity band in $^{101}$Pd was
extended up to I=$51/2^-$ and the band structure was interpreted as an AMR band following the SCM analysis for this system. 
It was demonstrated that rotational properties of the band can be described using the shears mechanism with
neutrons aligned along the rotational axis, and the proton pair in the time-reversed orbits aligning towards the
rotational axis with increasing spin.

The PSM calculations for $^{101}$Pd have been performed with two sets deformation parameters, listed in Table \ref{tab1}.
The lower value of deformation has been adopted from the standard mass table of ref.~\cite{Moller1995} and 
also used in the number conserved CSM calculations \cite{101pd}. The higher deformation value has been
employed from the total Routhian surface (TRS) studies on $^{101}$Pd, obtained for higher rotational frequencies. We
shall demonstrate in the following
that lower deformation set reproduces the low-lying spectrum, whereas the higher deformation value is needed
to describe the high-spin characteristics.

The PSM calculated energies with two sets of deformation values along with the
observed energies are displayed in Fig.~(\ref{levelpd1}).
It is evident from the figure that PSM results for $\alpha=-1/2$ branch with $\epsilon=0.128$ are in
good agreement with the data up to
I=$27/2^-$ and above this spin value it is the second set with $\epsilon=0.165$ that depicts a better agreement.
It will be shown below that
in order to reproduce the observed bandcrossing features properly in $^{101}$Pd,
it is essential to employ the two sets of deformation parameters.

For the $\alpha=+1/2$
branch, a deviation of about 0.3 MeV is noted. This deviation is recognized from the lowest I=17/2$^-$ state of
the $\alpha=+1/2$ structure. The reason for  this overall shift in the energies of the $\alpha=+1/2$ component is not 
understood at this stage and more work needs to be performed. However, it needs to be pointed out that 
spin assignment for $\alpha=+1/2$ signature branch in the experimental work of ref. ~\cite{Zhou2011} is tentative. 

To analyze the intrinsic structures of the bands, the band diagrams of $^{101}$Pd are displayed in
Fig.~(\ref{bd_pd1}) for the two deformation sets. In the band
diagram, the projected energy of each intrinsic configuration is depicted as a function of the angular-momentum. These
projected bands, before the shell model diagonalization, have diabatic crossings and it is easy to decipher the
nature of the bandcrossings. It is noted from the band diagrams that that the projected bands from K=1/2 and -3/2 configurations
are almost degenerate for the low-spin states with both deformation values. It is expected that ground-state
band will be composed of these
two configurations. These states are crossed in the high-spin region by three-quasiparticle states, and the crossing
features are different for the two sets. For the lower deformation value, shown in  the upper panel of Fig.~(\ref{bd_pd1}),
the three quaisparticle state with aligned two-protons crosses at I=27/2$^-$ and becomes favoured in energy. In the case
of larger deformation value, shown in the lower panel of Fig.~(\ref{bd_pd1}), the one-quasiparticle
and three-quasiparticle structures are almost degenerate for the spin states of I=27/2$^-$, 31/2$^-$ and 35/2$^-$, and, therefore,
the yrast band will be mixed for these states with the consequence that backbending is not expected. However,
the five-quasiparticle state becomes favoured in energy at I=47/2$^-$ with larger deformation value, whereas for the lower deformation
set, this crossing is delayed. This five-quasiparticle configuration has three neutrons in the N=5 shell and two-protons
in the N=4 shell. It is noted that five-quasiparticle state with one-neutron in the N=5 shell, two-neutrons in the N=4
shell and two-protons in the N=4 shell is almost degenerate with the three-quasiparticle configuration for the
intermediate spin states. 

The crossing for the $\alpha=+1/2$ member occurs at an earlier spin
value of I=17/2$^-$ for both deformation sets, and is due to the alignment of two-neutrons.
However, it is noted that neutron- and proton-aligned
configurations are very close in energy and it will be discussed later that bandcrossings are mixed.

The negative parity band structure in $^{103}$Pd was established as the AMR band through the lifetime measurements
\cite{103pdamr}. It was observed that B(E2) transitions drop with spin along the band, and also
the ratio of moment-of-ratio to B(E2) was found to be quite large. Theoretical calculations using covariant density functional
and SCM approaches were shown to be good agreement with the experimental data \cite{103pdamr}. However, the comparison of the bandcrossing
frequencies exhibited discrepancies \cite{103pdamr}. The calculations predicted almost simultaneous neutron and proton crosssings at
$\hbar \omega = 0.48$ MeV and 0.50 MeV, respectively. The observed bandcrossings are noted
to be at $\hbar\omega$=0.55 MeV and 0.64 MeV.

PSM calculated energies for the two signature branches of $^{103}$Pd and the known experimental band for the
$\alpha=-1/2$ are depicted in Fig.~(\ref{levelpd2}). The comparison of the
$\alpha=-1/2$ component shows an overall good
agreement. The deviation is noted to be about 100 keV for most of the spin states.
The $\alpha=+1/2$ states have not been observed for this
system, and PSM energies are provided in Fig.~(\ref{levelpd2}) for future comparisons.  

The band diagram of $^{103}$Pd, displayed in Fig.~(\ref{bd_pd2}), reveals that one-quasiparticle neutron configurations with
K=1/2 and -3/2 are favoured for low-spin states, and the yrast-state band will have dominant contributions from these
two configurations. It is noted from the figure that ground-state is crossed by a three-neutron state from the N=5 shell
at I=27/2$^-$ for the $\alpha=-1/2$ signature branch, and up to I=39/2$^-$ this three-quasiparticle state remains the yrast configuration.
The figure also shows that five-quasiparticle state having K=1/2, becomes yrast for $\alpha=-1/2$ at I=43/2$^-$ and above spin states. For the
$\alpha=+1/2$ branch, the first bandcrossing is noted at I=29/2$^-$ and the second crossing is seen at I=37/2$^-$.

High-spin states in $^{107}$Pd were studied using the heavy-ion fusion-evaporation reaction, ($^{18}O$+$^{96}Zr,)$
\cite{Pohl2682}, and the negative parity yrast band
was established up to I=$51/2^-$. The comparison of the experimental energies with the PSM calculations for the $\alpha=-1/2$ branch,
shown in Fig.~(\ref{levelpd2}), illustrates that PSM reproduces the data quite well. The corresponding
band diagram in  Fig.~(\ref{bd_pd2}), depicts
the intrinsic configurations of the bands in $^{107}$Pd. It is noted from the figure that three projected bands from K=1/2, -3/2 and
5/2 Nilsson configurations are almost degenerate from I=5/2$^-$ to 23/2$^-$ and the ground-state band, obtained after shell model diagonalization,
is expected to be a mixed state of the three configurations. The ground-state configuration is crossed by a three-neutron
state at I=27/2$^-$, and a five quasiparticle state, composed of three-neutrons and two-protons, become yrast at I=37/2$^-$.

The neutron-rich $^{109}$Pd and $^{111}$Pd were probed through fission fragments following the fusion reaction \cite{109pd} and the
high-spin properties of the two isotopes were populated. For  $^{109}$Pd, both signature branches of the yrast
negative parity band were identified, whereas for  $^{111}$Pd only the favoured branch was populated
in the experimental work \cite{109pd}. These
observed energies are compared with the PSM calculated energies in Fig.~(\ref{levelpd2}), and it
is noted that PSM approach reproduces the data quite well.
The band diagrams of the two isotopes, Fig.~~(\ref{bd_pd2}),  reveal that the first bandcrossing is due to the alignment of two-neutrons in the $1h_{11/2}$ subshell,
and the second crossing is due to further alignment of two-protons in the $1g_{9/2}$ subshell.

We shall now turn our discussion to  the five Cadmium isotopes studied in the present work. 
$^{103}$Cd has been investigated using the in-beam spectroscopic methods via the $^{72}$Ge($^{35}$Cl, p3n) reaction
\cite{103cdbt} and the negative parity yrast band has been populated up to I=47/2$^-$. It is observed
that this band depicts a regular rotational spectrum up to I=39/2$^-$, and above this spin irregular pattern of several parallel transitions
are observed, indicating a possible change of level structure at high-spin. TRS calculations  suggest that
the collective prolate shape obtained at low rotational frequencies, transitions into a non-collective oblate shape at higher
rotational frequencies \cite{103cdbt}. The comparison of the calculated TPSM energies with the experimental energies is shown in Fig.~(\ref{levelcd1})
and it is noted that PSM approach reproduces the data well up to I=31/2$^-$, but above this spin large deviations are noted.
The reason is quite clear as mean-field prolate deformation is kept fixed in the PSM calculations, and at higher spin shape transition to
non-collective oblate shape is predicted by the TRS calculations \cite{103cdbt}. This shape tranitions is neglected in the present
version of the PSM approach, and will need implementation of the generator coordinate method \cite{balian,hill53,gfiffin57,VALOR2000,RODRIG02,egido16}. 
The band diagram for $^{103}$Cd, shown in Fig.~(\ref{bd_cd2}), depicts
a bandcrossing at I=25/2$^-$ due to the alignment of two-neutrons in the N=4 shell.

The negative parity yrast band in $^{105}$Cd has been established to arise from the AMR mechanism
\cite{105cdamr}. The lifetime
measurements for this system have been performed using  the doppler shift attenuation technique. The deduced B(E2) transitions beyond I=23/2$^-$ decrease
with increasing spin, and the ratio of moment-of-inertia to B(E2) has a large value as expected for an AMR structure. SCM  calculations were performed and
it has been shown that data is reproduced well with the AMR configuration for valence neutron particles and proton holes \cite{105cdamr}. The PSM calculated energies
are compared with  the known experimental energies in Fig.~(\ref{levelcd1}) and it is evident that PSM reproduces the data reasonably well. There is a small deviation
of about 0.1 MeV for the highest observed spin, I=47/2$^-$ in the $\alpha=-1/2$ signature branch. For the $\alpha=+1/2$, the deviation is much higher, and
it needs to be added that the spin assignment for the I=37/2$^-$ is
tentative \cite{105cdamr}. The band diagram of $^{105}$Cd, presented in Fig.~(\ref{bd_cd2}), depicts an interesting crossing features. It is
noted that $\alpha=+1/2$ one-
quasiparticle state is crossed by a three-quasiparticle with two-neutrons in the N=4 shell at I=21/2$^-$. For the $\alpha=-1/2$, the crossing at I=27/2$^-$
is due to the alignment of two-neutrons in the N=5 shell. As a matter of fact, there are many three-quasiparticle states including one-neutron plus
two-protons in the N=4 shell, which are close in energy and it is expected that the yrast band after I=21/2$^-$ will have a mixed configuration. There has
been a detailed investigation regarding the nature of the bandcrossings in $^{105}$Cd and it has been shown that for a lower deformation value, protons
will align first, whereas for a larger deformation value the two-neutron alignment will occur before the alignment of two protons
\cite{reganjp}. It will be shown later, through the wavefunctions analysis, that the first bandcrossing
in $^{105}$Cd is due to a superposition of two-proton and two-neutron aligning configurations.

High-spin of the yrast negative parity band in $^{107}$Cd were investigated using the  
$^{94}$Zr($^{18}$O,5n) reaction \cite{107cd15}. The B(M1) probabilities were measured for five transitions
in the high-spin region and
they depict a decreasing trend with spin, which is a characteristic property of the MR band.
Tilted axis cranking (TAC)  analysis was also performed and it was shown that
high-spin states beginning from I=29/2$^-$ have a five-quasiparticle structure. Further, a backbend is observed at $\hbar \omega = 0.72$ MeV
and it was shown from the TAC calculations that it is due to the crossing of two five-quasiparticle configurations. 

The PSM calculated
energies for $^{107}$Cd are compared with the known data in Fig.~(\ref{levelcd1}), and it is noted that PSM reproduces the data reasonably well. The major discrepancy
of about 0.6 MeV is seen for the highest known spin value of I=49/2$^-$. The band diagram, plotted in Fig.~(\ref{bd_cd2}), depicts interesting crossing
features. The one-quasineutron band at low-spin is crossed by a three-quasiparticle band at I=17/2$^-$. The
three-quasiparticle band is then crossed by a five-quasiparticle band at I=31/2$^-$. This five-quasiparticle structure has one-neutron in the
N=5 shell, two-neutrons in the N=4 shell, and two-protons in the N=4 shell. It is also seen that this 5-qp is crossed by another
5qp state with three-neutrons in the N=5 shell and two-protons in the N=4 shell at I=47/2$^-$. This crossing obtained in the PSM calculations is
somewhat higher than observed in the data, which is around I=41/2$^-$. The reason for this discrepancy could be due to modification
of the men-field potential at high-spin. Actually, TAC calculations predict a transition from oblate to prolate shape at high-spin
in $^{107}$Cd.

Shears bands, both AMR and MR, band structures have been observed in $^{109}$Cd \cite{chiara20}. The yrast-band
with one-neutron in low-$\Omega$ orbit, and two-neutrons in high-K time-reversed orbits, has been shown to have the AMR structure. The excited
band with one-neutron and two-protons aligned along the rotational axis, has been demonstrated as having the MR characteristics. The experimental
evidences of AMR and MR band structures have been corroborated with TAC and SCM theoretical calculations \cite{chiara20,cdSCM}.
The PSM calculated energies for $^{109}$Cd are compared with the known experimental energies in Fig.~(\ref{levelcd2}). It is noted from the figure that
PSM reproduces the data quite well for low- and medium spin states, however, for very high-spin states, deviations are noted. The
band diagram for $^{109}$Cd, displayed in Fig.~(\ref{bd_cd2}), shows the alignment of two-neutrons in the N=5 shell at I=23/2$^-$, and then further
alignment of two-protons in the N=4 shell at I=39/2$^-$.

Band structures in $^{111}$Cd have been investigated in several experimental works \cite{regan1885,Cd94}. The predictions of the PSM calculations are compared with
the known energies in Fig.~(\ref{levelcd2}),
and it is evident from the figure that PSM reproduces the data quite well for the $\alpha=-1/2$ signature branch. The band diagram,
shown in Fig.~(\ref{bd_cd2}), depicts the bandcrossings at I=23/2$^-$ and 43/2$^-$.

\subsection{ Spin versus rotational frequency }

To investigate the bandcrossing phenomena in detail, the behaviour of angular-momentum, $I$, as a function of
rotational frequency ($\hbar \omega$) is presented in Figs.~(\ref{ixpd}) and (\ref{ixcd}) for the two isotopic chains of Pd- and Cd-nuclei.
The plot in Fig.~(\ref{ixpd}) for $^{101}$Pd, obtained from the measured values, depicts an upbend at around $\hbar \omega =0.45$MeV
and also a backbend at about $\hbar \omega = 0.6$ MeV. PSM calculations with $\beta=0.128$ reproduces the upbend at
$\hbar \omega = 0.45$MeV, and is due to the alignment of two-protons as is evident from the band diagram of Fig.~(\ref{bd_pd1}). However, the
backbending observed at $\hbar \omega=0.6$ MeV is not seen in the PSM with this deformation value. The PSM calculations with
the larger deformation value are shown in the inset of Fig.~(\ref{ixpd}) and it is noted that upbend at $\hbar \omega =0.45$ MeV
is now absent, but the bandbend at $\hbar \omega =0.6$ MeV is reproduced. We would like to mention that in the number-projected
CSM calculations performed in ref.~\cite{101pd} with the smaller deformation value, the bandbend at $\hbar \omega=0.6$ MeV is also absent.
It can, therefore, be inferred from the bandcrossing analysis that $^{101}$Pd undegoes a shape transition in the high-spin region.

For the $\alpha=+1/2$ signature component, the crossing features are somewhat different from the $\alpha=-1/2$ branch.
A slight backbend is observed at about $\hbar \omega = 0.45$ MeV, and the PSM calculations also depict a backbend but
with a weaker interaction between the ground-state and the aligned band as compared to the experimental data. The weaker interaction
in the PSM calculations is deduced from the larger curvature of $I$ verses $\hbar \omega$ curve.

The PSM calculated $\alpha=-1/2$ signature branch of $^{103}$Pd in Fig.~(\ref{ixpd}) depicts an upbend at around $\hbar \omega = 0.55$ MeV, whereas the
$\alpha=+1/2$ branch shows a backbend at $\hbar \omega = 0.57$ MeV and an upbend at $\hbar \omega=0.67$ MeV. The results corresponding 
to the experimental data also has an upbend at $\hbar \omega =0.57$ MeV for the $\alpha=-1/2$ branch.

For $^{107}$Pd, the observed backbend at about $\hbar \omega = 0.42$ MeV for $\alpha=-1/2$ branch is well
reproduced by the PSM calculations. The PSM calculations show backbend at a slightly higher rotational
frequency for the $\alpha=+1/2$ signature partner branch, and another backbend at around $\hbar \omega =  0.68$ MeV.
$^{109}$Pd depicts a similar bandcrossing features as that of $^{107}$Pd. In the case of $^{111}$Pd, PSM calculations
show a backbend at a similar rotational frequency as that $^{107}$Pd and  $^{109}$Pd, but the experimental data is not
available for the high-spin states to draw a comparison.

For the studied Cd-isotopes, the results of $I$ verses $\hbar \omega$ are displayed in Fig.~(\ref{ixcd}), and it is
evident that PSM calculations are in good agreement with the experimental data. The backbending is
observed for $^{103}$Cd, $^{109}$Cd and $^{111}$Cd isotopes at about the same rotational frequency for the
$\alpha=-1/2$ branch, and PSM
approach reproduces this bandbending feature. For $^{105}$Cd, an upbend is observed which is also reproduced
reasonably well by the PSM calculations. $^{107}$Cd depicts an upbend at a much higher rotational frequency
of $\hbar \omega = 0.72$ MeV. For the $\alpha$=+1/2 branch, the backbending is observed in all the studied
isotopes except for $^{107}$Cd.

Further, the bandcrossings for odd-neutron Pd- and Cd- isotopes have a complex structure and cannot be solely attributed
to either alignment of two-protons or two-neutrons. In order to illustrate the complex nature of the bandcrossings
for the studied nuclides, the wavefunctions of two representative cases of $^{101}$Pd and $^{105}$Cd are depicted
in Fig.~(\ref{wf}). The displayed wavefunction amplitudes have been orthogonalized following the procedure outlined in
Refs.~\cite{WANG2020,np}.
It is noted from this figure that for $^{101}$Pd, the yrast band is initially a one-quasiparticle state
and is crossed by a three-quasiparticle state, having two-neutrons in the N=4 shell, at around I=17/2$^-$. Although this
state is dominant, but there are many other configurations contributing significantly after the bandcrossing. For instance, the
wavefuction has finite contribution from the three quasiparticle state having alignment of two-protons. In the high-spin
region, the wavefunction is dominating by several five-quasiparticle configurations.

The wavefunction for $^{105}$Cd, shown in the lower panel of Fig.~(\ref{wf}), also depicts a complex structure after the bandcrossing
at I=19/2$^-$. It is to be noted that after the alignment, the favoured and unfavoured signature branches have different
structures. For the unfavoured branch, the crossing occurs first and has dominant three neutron configuration
with one neutron in the N=5 shell
and two aligning neutrons in the N=4 shell. In the case of favoured signature branch, the crossing occurs later, and after the
crossing the dominant configurations is a three neutron state with all the neutrons in the N=5 shell.

\subsection{Electric quadrupole transition probabilities}

The important characteristic feature of the AMR structure is revealed through the lifetime measurements \cite{101pdsuga2,103pdamr,109cd}. It
has been shown using the SCM that B(E2) transitions should decrease with increasing spin along the band, and also
ratio of moment-of-inertia to B(E2) should have a large value. The latter feature is important as it distinguishes AMR from the
phenomenon of terminating bands. For terminating bands, B(E2) as well as moment-of-inertia decreases, but in the case of AMR
only B(E2) drops and the moment-of-inertia stays constant along the band. For the MR structures, the B(M1) transitions depict
a decreasing trend with spin, and the B(E2) values are almost zero.

%Lifetime measurements for the $1/2^- [550]$ band
%in $^{101}$Pd were
%performed \cite{101pdsuga2} and it was observed that B(E2) transitions decrease with spin and the
%ratio of moment-of-inertia to B(E2) tend to have a large value.

In Figs.~(\ref{be2pd}) and (\ref{be2cd}), BE(2) transition probabilities are depicted for the two isotopic chains
of Palladium and Cadmium, respectively. The transition probabilities have been calculated with the expressions
given in Section~\ref{Sect.02} and the effective charges of
$e_p=1.5e$ and $e_n=0.5e$ for protons and neutrons, respectively. In most of the cases, it is noted that PSM calculated
B(E2) transitions decrease with spin. The detailed experimental transitions have been measured only
for three nuclides of $^{101}$Pd, $^{103}$Pd  and $^{105}$Cd, and it is evident from the results
that PSM approach provides a good description of the known data. It is to be noted that for $^{107}$Cd, the
BE(2) transitions are almost zero as expected since the band has MR character.
A few data points are
also available for $^{109}$Cd, and PSM values for the last two spin values are in good
agreement with the data. For the spin value of I=35/2$^-$, the known B(E2) transition probability
is higher than the PSM predicted value, and the reason for this discrepancy is not
clear.

The ratios of moment-of-inertia to B(E2) transitions are plotted in Figs.~(\ref{j2pd}) and (\ref{j2cd})
for the two studied isotopic chains. The ratios in most of the cases remain constant,
however, there are also some points where the ratios increase and decrease. These
changes are due to the bandcrossings as moment of inertia being second derivative is
very sensitive to any changes in the structure.

We have also evaluated the BM(1) transitions with the effective g-factors of : $g_s(\nu)=-3.826\times0.70, g_l(\nu)=0$
$(g_s(\pi)=5.586\times 0.70, g_l(\pi)=1)$ for neutron (protons), and are displayed in Fig.~(\ref{bm1}) for all the ten nuclides
studied in the present work. As expected, BM(1) transitions are almost zero for all the cases, except for
$^{107}$Cd, which has the MR character. It is evident from the figure that calculated BM(1) transitions
are in good agreement with the known transitions for this system.

For $^{109}$Cd, one of the excited bands has been shown to have the MR properties \cite{Cd94,chiara20}.
In the PSM calculations, the excited bands are quite mixed and it is difficult to identify the observed
MR band structure. However, it is known that MR band has the $1\nu 2\pi$ configuration and we have performed
the PSM calculations with the basis space of $1\nu$ and $1\nu 2\pi$ configurations only. The calculated
BE(2) and B(M1) transitions are depicted in Fig.~(\ref{bebm}) and it is evident from the figure that BM(1)
transitions are in excellent agreement with the known transitions. On the other hand, the BE(2)
transitions are almost zero as expected for an MR band structure.

%===========================SCM Discussion====================================

The electric quadrupole transition probabilities, B(E2), for the negative parity AMR  bands
in $ ^{103,105,109,111} $Cd and $ ^{101,103,107} $Pd isotopes have also been calculated using the semi-classical
particle-rotor model. In these calculations, $ \pi(g_{9/2})^{-4} \otimes \nu(h_{11/2})(d_{5/2}/g_{7/2})^{2} $ and
$ \pi(g_{9/2})^{-2} \otimes \nu(h_{11/2})(d_{5/2}/g_{7/2})^{2} $ quasiparticle configurations have been considered for Pd and Cd isotopes,
respectively. Experimentally, the antimagnetic rotational character of the yrast negative parity band beyond alignment has been established conclusively in $^{105,109} $Cd from lifetime measurements \cite{105cdamr,chiara20}. The SCM calculation nicely reproduce
the experimentally measured B(E2) values \cite{105cdamr,109cd}. The present calculations, shown in Fig.~(\ref{be2cd}), are
also found to be in agreement with the previously reported values. Although the band-head spins differ in the two Cd isotopes as
they exhibit rotational alignment at different frequencies.

As already mentioned, the first AMR band in Pd-isotopes was identified in $ ^{101} $Pd \cite{101pdsuga,101pdsuga2} and
unlike odd-Cd
isotopes, here the lower spin part of the $ \nu h_{11/2} $ band was proposed as the AMR band \cite{101pdsuga2}. But, in a
subsequent study, it was contradicted and the AMR character of $ \nu h_{11/2} $ band was proposed within $31/2^{-} $ to $ 43/2^{-} $ spin range \cite{Singh_2017}. Later, AMR motion was also found in $ ^{103} $Pd, where the measured B(E2) values were found in good agreement with SCM,
as well as with TAC-CDFT theoretical estimates at higher spin \cite{103pdamr}. However, the latest study on $ ^{103} $Pd suggests that
the angular momentum in this band arises from a combined contribution of antimagnetic rotation, collective rotation, and gradual
neutron alignment \cite{Deo2021Antimagnetic}. The present SCM calculation, however, reasonably reproduce the experimentally deduced B(E2) values as depicted in Fig.~(\ref{be2pd}).

It is interesting to note that the calculated B(E2) transitions of $ ^{103,111} $Cd and $ ^{107} $Pd are also found close to those
measured
experimentally for the well established AMR bands in $ ^{105,109} $Cd and $ ^{101,103} $Pd (\textit{see}, Figs.~(\ref{be2pd}) and (\ref{be2cd})).
Thus, the higher spin part, after alignment, of the negative parity yrast band in $ ^{103,111} $Cd and $ ^{107} $Pd are the
possible candidates for AMR. The SCM results are also compared with those obtained from the newly developed extended PSM approach
in this work, and it is  evident from Figs.~(\ref{be2pd}) and (\ref{be2cd}) that the results from the two approaches are in good
agreement.

%However, no such suitable negative parity E2 cascade has been found in $ ^{107} $Cd which could be a possible candidate for AMR band. Therefore, it opens up an avenue for further experimental research.

\section{Summary and conclusions}\label{sum}

The main objective of the present work has been to develop a microscopic approach to investigate the properties of MR and AMR
band structures. These structures have been observed in the vicinity of the closed shells and the characteristic feature of the AMR and MR structures
is that they have a regular rotational spectrum, but the electric quadrupole transitions are quite weak as compared to a typical deformed
nucleus. These new phenomena have been discussed using the shears mechanism with maximum contribution to angular-momentum originating from a 
few valence protons and neutrons, and the core contribution is very small.

In the present work, an extended PSM approach has been developed with quasiparticle excitations considered from
two major oscillator shells. In the original PSM version,
the excitations were restricted to one major shell only, although the vacuum configuration is generated from the three shells. It has been
demonstrated in several studies that PSM approach provides an excellent description of the properties of the deformed nuclei
in the neighbourhood of the yrast line \cite{SLATHIA201539,CHAUDHARY201553,LIU201111,PhysRevC.85.054307}.
The high-spin states along the yrast line are mostly generated by considering the quasiparticle excitations from the intruder orbital, which is a
part of the one major shell chosen in the PSM basis space. However, to describe the non-yrast band structures, the quasiparticle
excitations from the other
major shells also needed to be included in the basis space. We have generalized the PSM approach to include the quasiparticle excitations
from two major oscillator shells, and have also significantly enlarged the rank of the quasiparticle space. In most of the
earlier studies, a maximum
of three quasiparticle excitations were considered for odd-mass nuclei, and in the present study we have included up to five quasiparticle
configurations in the basis space.

As a first application of the generalized PSM approach, we have investigated odd-mass Pd- and Cd-isotopes. These nuclei are situated in the vicinity of the
N=Z=50 closed shell and in many of these nuclei, MR and AMR band structures have been observed. The high-spin states in these bands are
mostly generated by the angular-momentum of the valence neutrons and protons. It has
been demonstrated that PSM provided a reasonable description of most of the observed properties of the studied ten nuclides. The excitation
energies for the favoured $\alpha=-1/2$ branch has been shown to be reproduced within 100 keV in almost all the cases. However, for the
unfavored $\alpha=+1/2$ branch, some discrepancies have been noted. But the data is quite limited for this branch, and more data
is needed to have a better assessment of the accuracy of the predicted values.

A detailed investigation of the bandcrossings has been performed of the studied isotopes. For $^{101}$Pd, it has been shown that in order to
reproduce the observed bandcrossings, one at $\hbar \omega$ = 0.45 MeV and the other 0.6 MeV, two different deformation values are needed in the PSM
calculations. The first crossing is due to the alignment of two-protons, and the second crossing is because of the further alignment of two-neutrons.
In all other Pd-isotopes, the first crossing is due to the alignment of two-neutrons in the N=5 shell. 

For $^{103}$Cd, the first band crossing is due to the alignment of two-neutrons in the N=4 shell. In the case of $^{105}$Cd, the first crossing in the $\alpha=-1/2$
branch is due to the alignment of two-neutrons in the N=5 shell, however, for the $\alpha=+1/2$ branch, it is due to the alignment of two-neutrons
in the N=4 shell. For $^{107}$Cd, an interesting crossing between two five quasiparticle configurations is noted. In the cases of $^{109}$Cd and $^{111}$Cd,
the first crossing is due to the alignment of two-neutrons in the N=5 shell.

The most important characteristic feature for the existence of the MR and AMR band structures is revealed through the measurement of
electromagnetic transitions. It has been shown using the phenomenological model that in the case of AMR, the BE(2) transitions should drop
with increasing spin and for the case of MR these should be vanishing. It has been shown that BE(2) transitions, indeed, drop for all the
cases except for $^{107}$Cd for which they are almost zero as it has an MR structure. On the other hand, BM(1) transitions are quite
opposite with almost
negligible for the AMR, and having the decreasing trend with spin for the MR case. We have also evaluated the ratios
of $\mathcal{J}^{(2)}$ and B(E2) and it has
been shown that they have almost constant values for most of the cases.

In conclusion, it has been shown that the extended PSM approach developed in the present work has provided a good description of the  magnetic and antimagnetic
rotational band structures observed in Pd- ad Cd -isotopes. This has been the first application of the generalized PSM approach, and in future we are
planning to investigate the MR and MR band structure in the positive parity band structures of these nuclides, and also explore the Pb-region, where these
band structures were first identified.

\section*{Acknowledgements}

N.N. acknowledges the Department of Science and
Technology (Government of India) for the award of INSPIRE fellowship under Sanction No. DST/INSPIRE
Fellowship/[IF200508].
S.C. gratefully acknowledges financial support from IEM, Kolkata for establishing the \textit{Aage Bohr Laboratory for Nuclear Science} (BoNuS Lab) under the grant-in-aid project numbers: IEMT(S)/2025/R/04-G28; IEM(S)/2025/S/08-G63. 

\bibliographystyle{apsrev4-2}
\bibliography{references}
\end{document}